\documentclass[twocolumn]{aastex63}

\graphicspath{{./}{figures/}}
\usepackage{pifont}
\usepackage{multirow}
\usepackage{amsmath}
\usepackage{color}
\usepackage{graphicx}
\usepackage{subfigure}
\usepackage{boxedminipage}
\usepackage{threeparttable}
\usepackage{cases}
\usepackage{CJKutf8}
\usepackage{lineno}

\shorttitle{GRB~230307A}
\shortauthors{Wang et al.}

\begin{document}
\begin{CJK*}{UTF8}{gbsn}

\title{What Powered the Kilonova-Like Emission After GRB~230307A in the Framework of a Neutron Star-White Dwarf Merger?}

\author[0000-0002-9738-1238]{Xiangyu Ivy Wang (王翔煜)}
\affiliation{School of Astronomy and Space Science, Nanjing University, Nanjing 210093, China}

\author[0000-0002-1067-1911]{Yun-Wei Yu (俞云伟)}
\affiliation{Institute of Astrophysics, Central China Normal University, Wuhan 430079, China; \url{yuyw@ccnu.edu.cn}}
\affiliation{Key Laboratory of Quark and Lepton Physics (Central China Normal University), Ministry of Education, Wuhan 430079, China}

\author[0000-0002-9037-8642]{Jia Ren (任佳)}
\affiliation{School of Astronomy and Space Science, Nanjing University, Nanjing 210093, China}

\author[0000-0002-5485-5042]{Jun Yang (杨俊)}
\affiliation{School of Astronomy and Space Science, Nanjing University, Nanjing 210093, China}

\author[0000-0002-6189-8307]{Ze-Cheng Zou (邹泽城)}
\affiliation{School of Astronomy and Space Science, Nanjing University, Nanjing 210093, China}

\author[0000-0002-9195-4904]{Jin-Ping Zhu (朱锦平)}
\affiliation{School of Physics and Astronomy, Monash University, Clayton Victoria 3800, Australia}
\affiliation{OzGrav: The ARC Centre of Excellence for Gravitational Wave Discovery, Clayton Victoria 3800, Australia
}

\begin{abstract}
The second brightest gamma-ray burst, GRB~230307A (with a duration $T_{90} \sim$ 40 s), exhibited characteristics indicative of a magnetar engine during the prompt emission phase. Notably, a suspected kilonova was identified in its follow-up optical and infrared observations. Here we propose that the origin of GRB~230307A is a neutron star-white dwarf (NS-WD) merger, as this could naturally interpret the long duration and the large physical offset from the center of its host galaxy. In the framework of such a NS-WD merger event, the late-time kilonova-like emission is very likely to be powered by the spin-down of the magnetar and the radioactive decay of $^{56}$Ni, rather than by the decay of $r$-process elements as these heavy elements may not be easy to be synthesized in a NS-WD merger. It is demonstrated that the above scenario can be supported by our fit to the late-time observational data, where a mass of ${\sim}10^{-3} \ \rm M_{\odot}$ $^{56}$Ni is involved in the ejecta of a mass of ${\sim}0.1 \ \rm M_{\odot}$. Particularly, the magnetar parameters required by the fit are consistent with those derived from the early X-ray observation.

\end{abstract}


\section{Introduction} \label{sec:intro}
In the past decades, it has been widely accepted that long GRBs are produced by the collapse of massive stars and short GRBs originate from compact binary mergers like mergers of double neutron stars (DNSs) and of a neutron star and a black hole \citep[NS-BH;][]{Eichler1989Natur, Narayan1992ApJ, Meszaros1992MNRAS}. The smoking-gun evidence to support such a common view is (i) the discovery of Type Ib/c supernovae emerging from the afterglow of some long GRBs \citep{Galama1998Natur, Bloom1999Natur, Hjorth2003Natur, Stanek2003ApJ, Woosley2006ARA&A} and (ii) the detection of short GRB~170817A \citep{Goldstein2017ApJ...848L..14G, Savchenko2017ApJ} in counterpart with the gravitational wave event GW170817 \citep{Abbott2017PhRvL}. In addition, accompanying the GW radiation, it is expected that DNSs and NS-BH mergers can simultaneously eject %outwards 
an amount of $\sim10^{-4}-10^{-2}M_{\odot}$ neutron-rich material \citep{Hotokezaka2013PhRvD, Radice2018ApJ}, which can lead to effective nucleosynthesis of heavy elements through rapid neutron-capture processes ($r$-processes). Then, the radioactive decay of these $r$-process elements can heat the merger ejecta and lead to an ultraviolet-optical-infrared transient emission, namely ``kilonova" \citep{Lilixin1998ApJ, Freiburghaus1999ApJ, Metzger2010MNRAS, Metzger2017LRR}. Such a kilonova model could explain the optical transient AT~2017gfo following the GRB~170817A/GW170817 event \citep{Abbott2017ApJ...848L..13A, Smartt2017Natur, Arcavi2017Natur, Coulter2017science, Pian2017Natur, Goldstein2017ApJ...848L..14G}, in despite of some difficulties \citep{Li&Yu2018ApJ}. 

Recently, suspected kilonova candidates have been surprisingly discovered from some GRBs of duration much longer than 2 seconds, which is a traditional boundary between long and short GRBs, e.g., GRB~060614 with $T_{90} \sim 100 \ \rm s$, GRB~211211A and GRB~230307A with $T_{90} \sim 40 \ \rm s$ \citep{Della2006Natur, Yangbin2015NatCo, Rastinejad2022Natur, Troja2022Natur, Yangjun2022Natur, Jinpingzhu2022ApJ, Wangyun2023ApJ, Levan2023, sunhui2023arXiv, Dichiara2023ApJ, Gillanders2023arXiv, yangyuhan2023arXiv, Mengyanzhi2023arXiv}. These facts challenge the traditional classification of GRBs, indicating that some long GRBs can be produced by compact binary mergers. Such a merger origin of GRB~230307A is further supported by its large offset of the location from its host galaxy center \citep[$\sim$ 36.60 kpc;][]{Levan2023, sunhui2023arXiv, Daicuiyuan2023arXiv}, as a compact binary can often travel far from its host galaxy before the merger due to the high kick velocity of the NS component in the binary \citep{Kalogera1998ApJ}. In principle, a long duration is not completely impossible to be produced by a merger event. For example, if a DNS merger has a high mass ratio $q \gtrsim 1.2$ and high total mass, or a NS-BH merger has a low mass ratio $q \lesssim 3$ and a high pre-merger BH spin, then the merger product (i.e., a BH) could be surrounded by a magnetically arrested massive ($ \sim 0.1 M_{\odot}$) disk, which could extend the energy release of the BH through an accretion tail of $\dot{M} \propto t^{-2} $ \citep{Ore2023arXivb, Ore2023ApJ}.

However, for GRB~230307A, the broad-band (0.5--6000 keV) observations during its prompt emission revealed that the central engine of this GRB is probably a rapidly rotating and highly-magnetized remnant NS \citep{sunhui2023arXiv}. First, its soft X-ray emission in the 0.5--4.0~keV energy range monitored by Lobster Eye Imager for Astronomy (LEIA) shows a different spectrum from the energy band above 15 keV monitored by the Gravitational wave high-energy Electromagnetic Counterpart All-sky Monitor (GECAM). 
Second, the late decay slope of soft X-ray emission does not conform to the curvature effect. Finally, the soft X-ray light curve is basically consistent with the shape of an initial plateau followed by a steep decay, which has been usually regarded as an observational signature of a spinning-down millisecond magnetar. In summary, the NS-BH merger origin of GRB~230307A can be ruled out. On other hand, although such a remnant mangetar can in principle be produced by a DNS merger, this scenario is still challenged by the long duration of the prompt emission. 
Furthermore, the existence of the remnant magnetar can also make the DNS merger scenario difficult to explain the $>$ 3 $\mu$m spectroscopic features of GRB~230307A \citep{Gillanders2023arXiv}, since the long-term neutrino emission from the magnetar can effectively suppress the synthesis of lanthanide elements \citep{Martin2015ApJ}.

In comparison, because the density of white dwarfs (WDs) is much lower than that of NSs, the free-fall timescale of a NS-WD merger can be much longer than that of a DNSs or a NS-BH merger, which may provide a more natural explanation for the origin of GRB~230307A. In the NS-WD merger scenario, a stable remnant magnetar is very probably produced because of the relatively low mass of the remnant \citep{Bartos2013CQGra, zhangbing2018pgrb.book}. However, the problem in this case is how the kilonova-like transient emission can be generated after the merger, since the ejecta from the NS-WD merger could not be an ideal place for synthesizing $r$-process elements \citep{Metzger2012MNRAS, Zenati2019MNRAS, fernndez2019MNRAS, Kaltenborn2022arXiv220913061K}. Instead, a small amount of $^{56}$Ni can be expected to form \citep{Alexey2022MNRAS, Zhongshuqing2023ApJ}, the radioactive decay of which can contribute to the thermal emission of ejecta. Besides, it needs to be noticed that the spin-down of the remnant magnetar could power the ejecta emission significantly, as previously suggested for other magnetar-driven transient phenomena \citep{Kasen2010ApJ, Inserra2013ApJ, Yunwei_yu2013ApJ, Metzger2014MNRAS, Yu2015ApJ, ShunkeAi2022MNRAS}.

Therefore, the purpose of this {\it Letter} is to test whether the parameters of magnetar constrained from the early X-ray emission of GRB~230307A can further explain the multi-wavelength afterglow emission plus kilonova-like emission in the framework of a NS-WD merger\footnote{The applicability of the magnetar engine model on the optical emission of GRB~230307A had been previously investigated by \cite{yangyuhan2023arXiv}, where, however, the constraint from the early X-ray emission and the radioactive decays of $^{56}$Ni were not taken into account. Additionally, the bolometric data used to constrain the model were derived from the properties of the black body component rather than the kilonova component, increasing some uncertainty.}. In Section \ref{sec:kilonova-like}, we introduce the model and present the fitting result. Conclusion and discussion is drawn in Section \ref{sec:discussion_and_conclusion}.

\section{Joint fit of Kilonova-like Component and Afterglow} \label{sec:kilonova-like}
\subsection{The Model}
As discussed above, the optical excess in the afterglow of GRB~230307A is considered to be powered by the combination of the spin-down of a newborn magnetar and the radioactive decay of $^{56}$Ni. Therefore, we first invoke the spin-down luminosity of the magnetar as usual as 
\begin{equation}
L_{\rm sd}(t)=L_{\rm sd}(0)\left(1+{t\over t_{\rm
sd}}\right)^{-2} \label{eq:Lsdt},
\end{equation}
where the spin-down is assumed to be dominated by the magnetic dipole radiation, $L_{\rm sd}(0)$ is the initial spin-down luminosity, and $t_{\rm
sd}$ is the spin-down timescale. In the following calculations, we fixed $t_{\rm sd}$ = 80 s according to the fit of the soft X-ray light curve during the prompt emission phase, where a smoothly broken power law is used with a break time of $ \sim$ 80 s \citep{sunhui2023arXiv}. Meanwhile, the radioactive power per unit mass of $^{56}$Ni and its daughter nucleus $^{57}$Co is given by
\begin{equation}
	\dot{q}_{{\rm r}}  = \epsilon_{\rm Ni} e^{-t / \tau_{\rm Ni}}+\epsilon_{\rm Co}
	\frac{e^{-t / \tau_{\rm Co}}-e^{-t / \tau_{\rm Ni}}} {1-\tau_{\rm Ni} / \tau_{\rm Co}},
	\label{eq:q_Ni}
\end{equation}
where $\epsilon_{\rm Ni}$ = $3.9 \times 10^{10}$ $\rm erg \ s^{-1} \ g^{-1}$ and $\epsilon_{\rm Co}$ = $6.8 \times 10^{9}$ $\rm erg \ s^{-1} \ g^{-1}$ are energy generation rate per unit mass \citep{Sutherland1984ApJ, Maeda2003ApJ}, and $\tau_{\rm Ni}$ = 8.8 days and $\tau_{\rm Co}$ = 111.3 days are the decay time.

A simplified radiation transfer model given by \cite{Kasen2010ApJ} and \cite{Metzger2017LRR} is adopted to describe the emission of the merger ejecta, where the ejecta is separated into $N$ layers with different velocities and densities. The density of each layer is determined by the following density profile of the ejecta \citep{Nagakura2014ApJ}:
\begin{equation}
\label{eq:density}
\rho_{\rm ej}(R,t)=\frac{(\delta - 3)M_{\rm ej}}{4\pi
R_{\max}^3}\left[\left(\frac{R_{\min}}{R_{\max}}\right)^{3 - \delta}-1\right]^{-1}\left({R\over
R_{\max}}\right)^{-\delta},
\end{equation}
where $M_{\rm ej}$ is the total mass of the ejecta, and $R_{\min}$ = $v_{\min}t$ ($R_{\max}$ = $v_{\max}t$) is the innermost (outermost) layer of the ejecta by introducing the minimum (maximum) velocity. $R = v_{\rm i}t$ is the radii of each layer assuming there is no dynamic evolution of each layer, where $v_{\rm i}$ is a constant velocity. The evolution of the thermal energy of the $i$th layer is determined by \citep{Yu_yunwei2018ApJ}
\begin{subnumcases}{}
    {dE_{i}\over dt} = \xi L_{\rm sd} + m_{{\rm {Ni}, i}} \dot{q}_{{\rm r},i}-{E_{i}\over R_{i}}{dR_{i}\over dt}-L_{i} , i=1 \label{eq:Ei_1},\\
    {dE_{i}\over dt} = m_{{\rm Ni}, i}\dot{q}_{{\rm r},i}-{E_{i}\over
    R_{i}}{dR_{i}\over dt}-L_{i}, i>1 \label{eq:Ei_2}.
 \end{subnumcases}
Here, the spin-down energy of the magnetar is assumed to be primarily absorbed by the ejecta at its bottom with an efficiency of $\xi$. $^{56}$Ni is considered to distribute the same as the ejecta does, where $m_{{\rm Ni}, i}$ represents the mass of $^{56}$Ni of each layer.
$L_{\rm i}$ is the observed luminosity of each layer, which can be calculated by
\begin{equation}
\begin{aligned}
&L_{i} = {E_{i}\over \max(t_{{\rm d},i},t_{{\rm lc},i})},
\end{aligned}
\end{equation}
where
\begin{equation}
t_{{\rm d},i} = {\frac{3\kappa}{4 \pi R_{i}c}} {\sum\limits_{j=i}^{n}}m_{j}
\end{equation}
is the radiation diffusion timescale for the photons from the $i$th layer to escape from the ejecta \citep{Arnett1982ApJ}, $\kappa$ is the opacity, and $t_{{\rm lc}, i} = R_{i}/c$ is the light crossing time. Then, the total luminosity of the merger ejecta is obtained by summing $L_{\rm i}$ up:
\begin{equation}
L_{\rm bol}={\sum\limits_{i=1}^{n}}L_{i}.
\end{equation}
Finally, the corresponding temperature of this thermal emission can be given by using the Stefan-Boltzmann law, and the specific luminosity for different colors can be derived from the Planck function by assuming a black-body spectrum for this emission. 

Generally speaking, the light curves given by our semi-analytical model can be consistent with the simulation results of \cite{Zenati2020MNRAS} and \cite{Kaltenborn2022arXiv220913061K} in some respects, although the specific frequency-dependence of the light curves could still be different. The differences can arise from the different distributions of the ejecta velocity and $^{56}$Ni mass and, in particular, the frequency-dependence of the opacity adopted in their simulations. Because of these complex factors, the spectra of the thermal emission of ejecta could deviate from the black body significantly. Nevertheless, due to the overlapping with the bright afterglow emission from the GRB jet, the thermal emission of ejecta can actually emerge only around its peak with a very limited data number. This makes it nearly impossible to constrain the above mentioned detailed properties of the ejecta, whereas the key features of the explosion system can still be inferred from the emission peak. Therefore, it is feasible and convenient to use the semi-analytic model to estimate the masses of the ejecta and $^{56}$Ni and the order of magnitude of the opacity for the primary emission band. Such a method has also been widely employed in literature for the fittings of supernova light curves \citep[e.g.,][]{Shigeyama1987Natur_SN1987, Pinto2000ApJ, Blinnikov2006A&A, Nicholl2016ApJ, Nicholl2017ApJ, Liangduan2021ApJ}.

\subsection{Fitting Results}

\begin{table}[]
\caption{The fitting parameters, prior bounds, and posterior medians at 1$\sigma$ credible interval.}
\renewcommand{\arraystretch}{1.5}
    \centering
    \begin{tabular}{lcc}

    \hline
    Parameter & Prior Bounds & Median $\&$ 1$\sigma$ C.I. \\
    \hline
    
    \multicolumn{3}{l}{\textbf{Afterglow}} \\

    log $E_{0}$ (erg) & [50.0, 60.0] & $51.11_{-0.49}^{+0.93}$ \\
    log $n_{0}$ (cm$^{-3}$) & [-6.0, 2.0] & $0.42_{-0.56}^{+1.10}$ \\
    log $\theta_{\rm c}$ (rad) & [-2.0, -0.5] & $-0.70_{-0.06}^{+0.05}$ \\
    log $\epsilon_{\rm e}$ & [-4.0, -0.2] & $-0.92_{-0.76}^{+0.51}$ \\
    log $\epsilon_{\rm B}$ & [-6.0, -0.2] & $-2.50_{-1.12}^{+0.39}$ \\
    $p$ & [2.01, 3.0] & $2.03_{-0.02}^{+0.02}$ \\
    log $\xi_{\rm N}$ & [-5.0, 0.0] & $-1.29_{-0.44}^{+0.74}$ \\
    
    \hline

    \multicolumn{3}{l}{\textbf{Kilonova-like Component}} \\
    
    $v_{\rm min}$ (c) & [0.001, 0.15] & $0.12_{-0.01}^{+0.01}$ \\
    $v_{\rm max}$ (c) & [0.18, 0.35] & $0.24_{-0.05}^{+0.06}$ \\
    $\delta$ & [1.0, 3.0] & $2.58_{-0.65}^{+0.33}$ \\
    log $\kappa$ ($\rm cm^2$ g$^{-1}$) & [-1.0, 0.0] & $-0.11_{-0.15}^{+0.08}$ \\
    log $M_{\rm ej}$ ($\rm M_{\odot}$) & [-2.0, -0.8] & $-0.93_{-0.19}^{+0.09}$ \\
    log $M_{\rm Ni}$ ($\rm M_{\odot}$) & [-5.0, -3.0] & $-3.32_{-0.04}^{+0.03}$ \\
    log $\xi L_{\rm sd}(0)$ (erg s$^{-1}$) & [45.0, 48.5] & $48.09_{-0.07}^{+0.09}$ \\

    \hline
    \end{tabular}
    \label{tab:fitting_para}
\end{table}

\begin{figure*}
    \centering
    \includegraphics[width=0.90\textwidth]{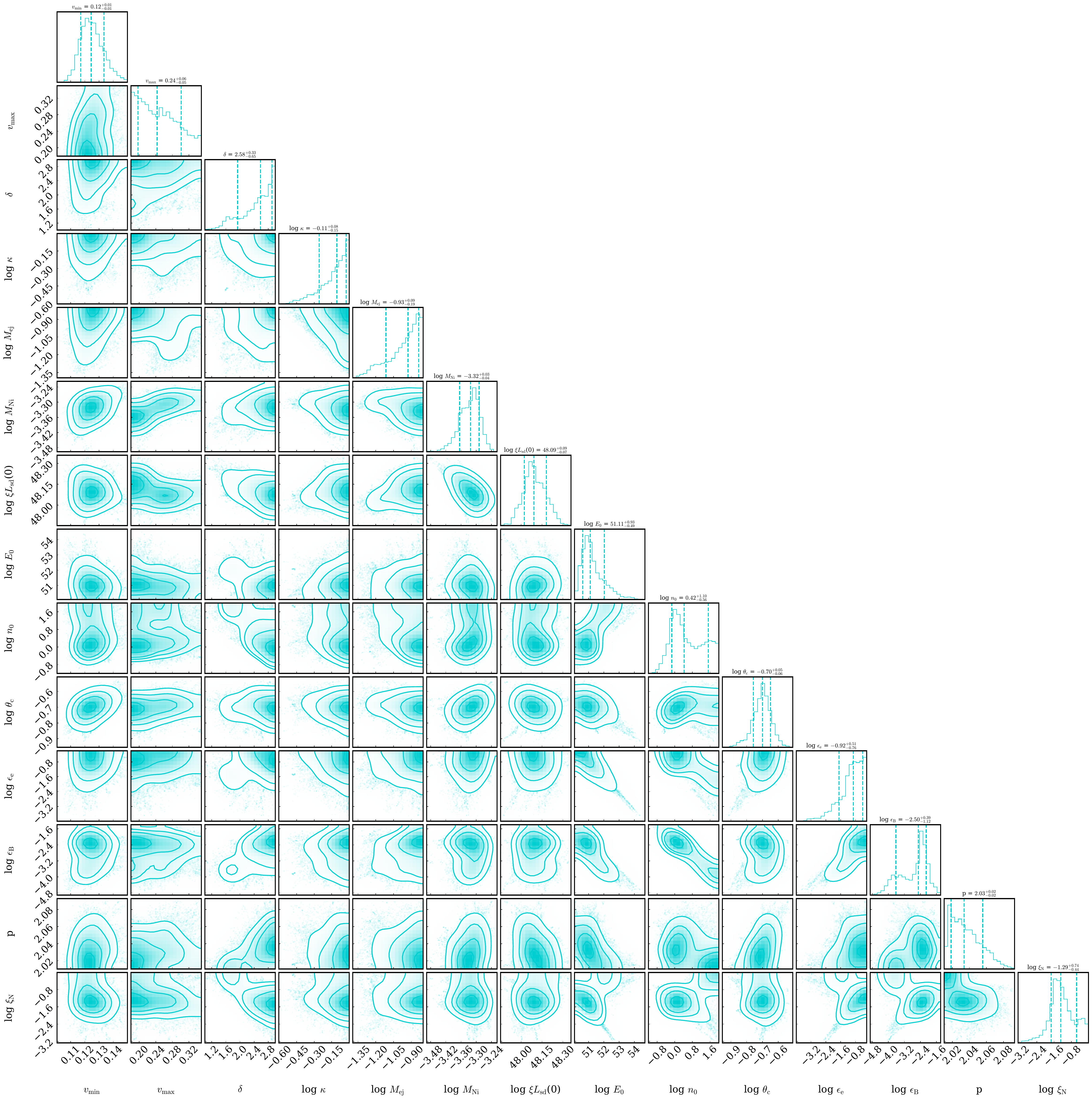}
    \caption{The corner plot of the posterior probability distributions of 14 parameters listed in Table \ref{tab:fitting_para} for the fitting of the multi-wavelength light curves.}
    \label{fig:corner}  
\end{figure*}

There are a total of 14 free parameters in our model. Besides 6 parameters for the merger ejecta and 1 parameter for the magnetar, we also need to invoke another 7 parameters to model the multi-wavelength afterglow light curves of the GRB, which is ascribed to the synchrotron radiation due to the shock interaction of a structured jet and the external medium \citep{Sari1998ApJ, Huangyongfeng1999MNRAS}. Here, a Gaussian structure $E(\theta) = E_{0} \rm{exp}(-\theta^2/2\theta_{\rm c}^2)$ is applied for the jet. The truncation angle is set to $\theta_{\rm w}$ = 4$\theta_{\rm c}$, and the viewing angle $\theta_{\rm v}$ is set to 0 rad in our fitting as same as those in \cite{yangyuhan2023arXiv}. The public python package \texttt{afterglowpy} \citep{Ryan2020ApJ} is used to reproduce the theoretical afterglow emission. The Markov Chain Monte Carlo (MCMC) method with the Python package \texttt{emcee} \citep{Foreman-Mackey2013PASP} is applied to fit the multi-wavelength light curves by minimizing $\chi^{2}$. We set 5000 walkers in a 56-dimensional Gaussian, and discard the first 2000 steps. The prior bounds, medians, and $1\sigma$ credible intervals for each parameter are given in Table \ref{tab:fitting_para} and the corner plot of the posterior probability distributions is shown in Figure~\ref{fig:corner}. 

\begin{figure}
    \centering
    \includegraphics[width=0.49\textwidth]{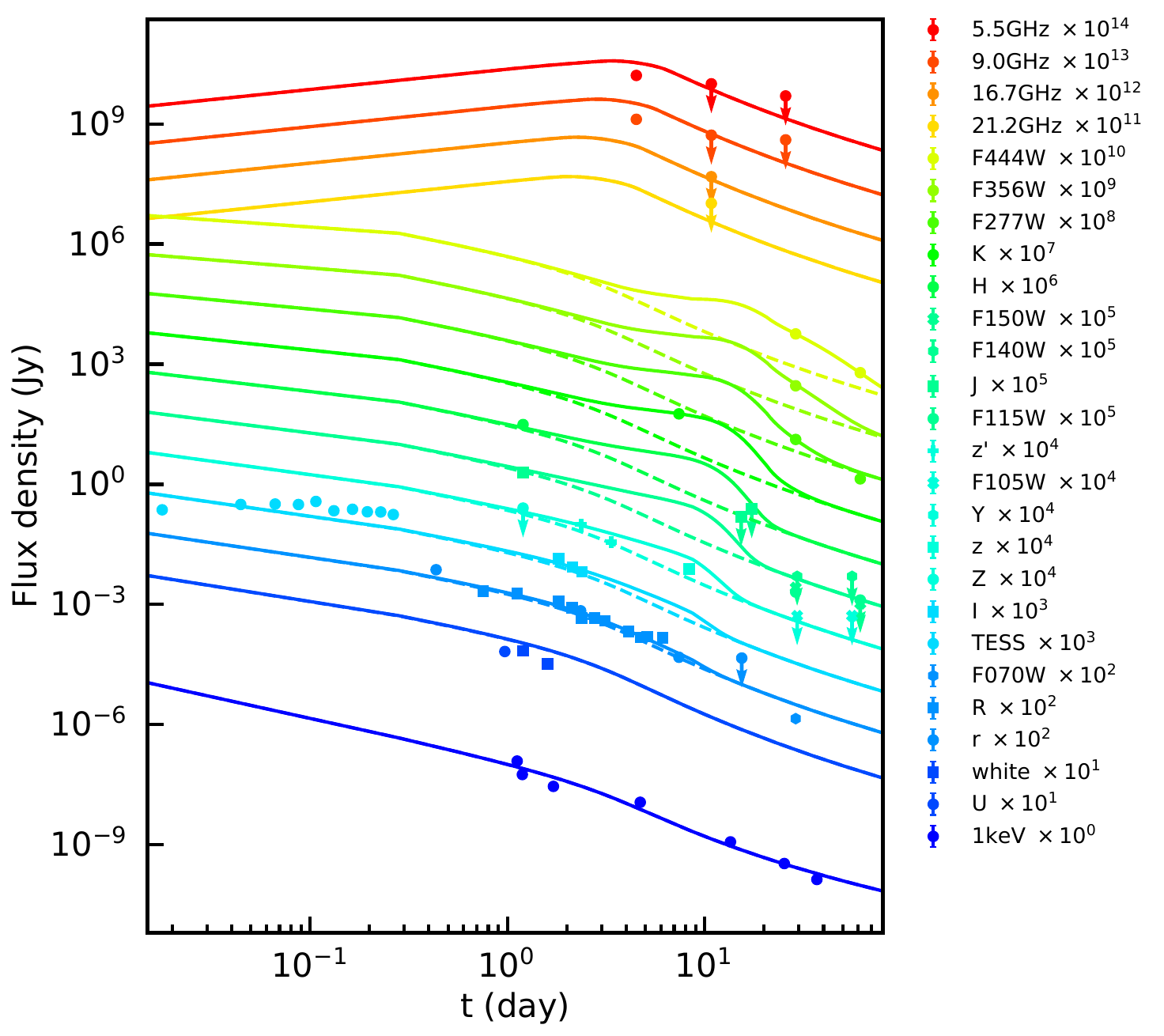}
    \caption{The multi-wavelength light curves. The solid line is the afterglow plus kilonova-like emission, while the dashed line is the afterglow-only emission. Both lines are plotted with the median parameters in our fitting. The data are collected from \cite{yangyuhan2023arXiv}.}
    \label{fig:afterglow}
\end{figure}

\begin{figure}
    \centering
    \includegraphics[width=0.45\textwidth]{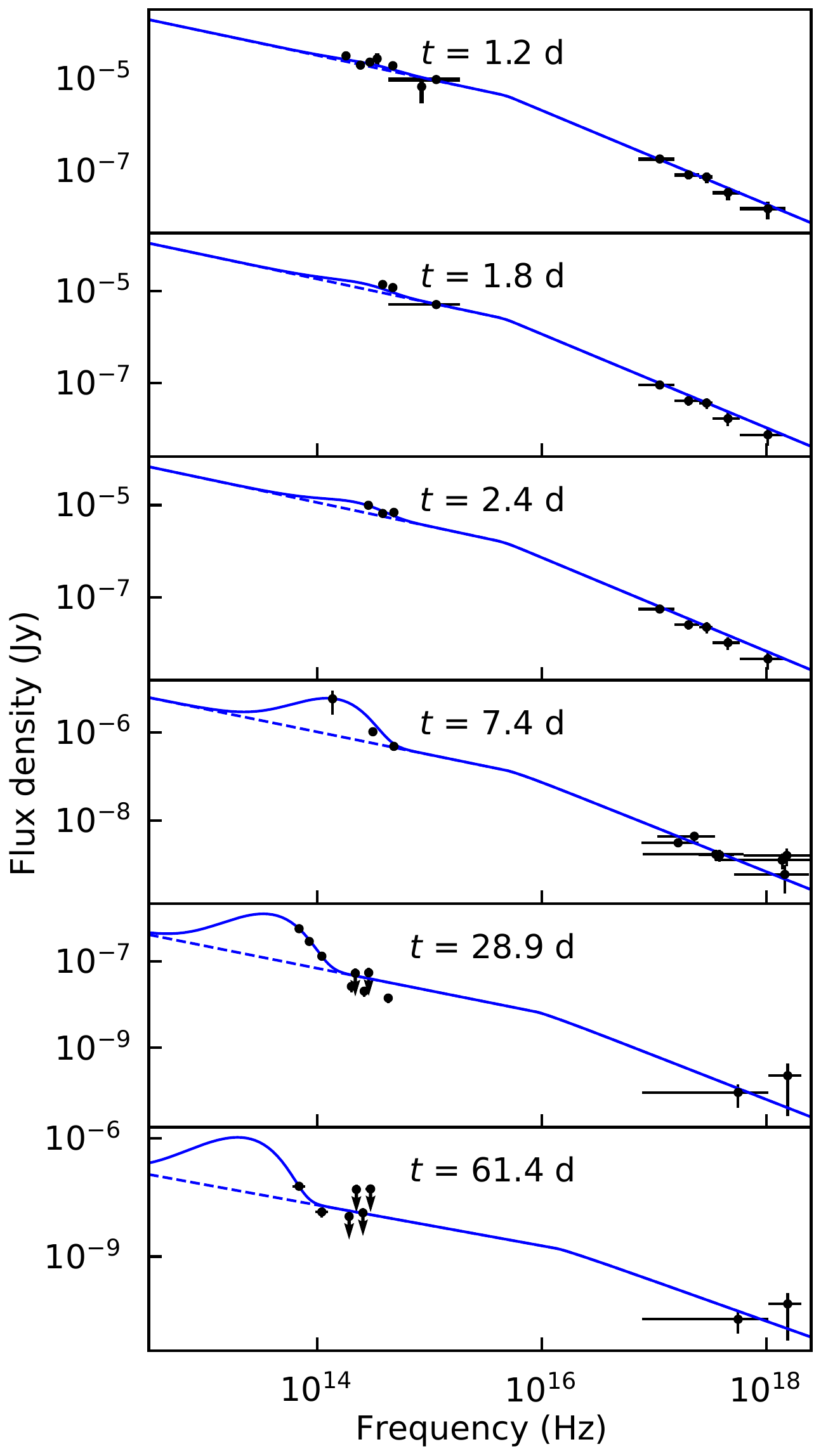}
    \caption{Spectral energy distributions. The black points are the unabsorbed spectra at different epochs from \cite{yangyuhan2023arXiv}. The solid and dashed lines are the afterglow plus kilonova-like emission and afterglow-only emission, respectively. }
    \label{fig:sed}  
\end{figure}

\begin{figure}
    \centering
    \includegraphics[width=0.49\textwidth]{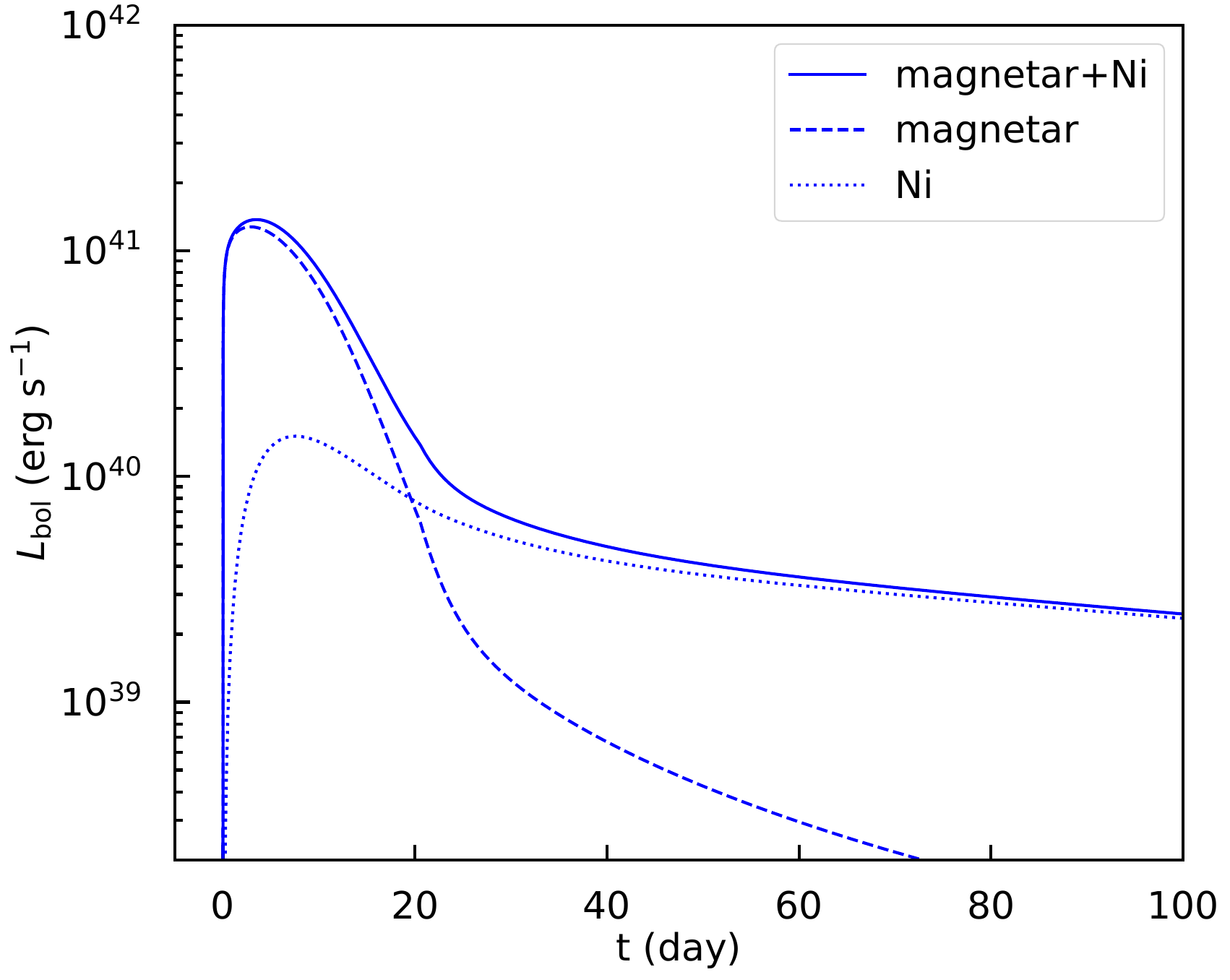}
    \caption{The bolometric light curves. The dashed and dotted lines represent the absorbed spin-down energy of the ejecta and the radioactive heating power from $^{56}$Ni, respectively. The solid line is the total energy of the two energy sources.}
    \label{fig:bolometric}
\end{figure}

With the median values of the model parameters, we plot the multi-wavelength light curves and the temporal evolving spectral energy distributions (SEDs) of the optical emission of GRB~230307A in Figures~\ref{fig:afterglow} and \ref{fig:sed}, respectively. The corresponding bolometric light curve is presented in Figure~\ref{fig:bolometric}. 
The bolometric light curve reaches its peak is in about 5 d in our fitting, similar to that in \cite{Kaltenborn2022arXiv220913061K} (5-8 d), but different from that in \cite{Zenati2020MNRAS} (10-20 d). The multi-wavelength light curves, spanning from the U-band to I-band, becomes redder over time in \cite{Zenati2020MNRAS} and \cite{Kaltenborn2022arXiv220913061K}, which can also be found from Figure \ref{fig:afterglow}. Furthermore, the obtained mass of $^{56}$Ni can be in the reasonable range predicted by the simulations. Figure \ref{fig:bolometric} shows that the radioactive power can dominate the radiation at the late stage, while the magnetar power dominates the first 20 days. 

To be specific, the initial absorbed spin-down energy from the magnetar is constrained to be $\xi L_{\rm sd}(0)$ = $1.23 \times 10^{48}$ erg s$^{-1}$ and the isotropic kinetic energy of the afterglow jet is $E_{0} = 1.29 \times 10^{51}$ erg. The total injected spin-down energy and the kinetic energy of the ejecta can be calculated as $E_{\rm total} \approx \xi L_{\rm sd}(0)$$\times t_{\rm sd}$ = $9.84 \times 10^{49}$ erg and $E_{\rm k} = \frac{1}{2}{\sum\limits_{i=1}^{n}}m_{\rm i}v_{\rm i}^{2}$ = $3.56 \times 10^{51}$ erg, respectively. Since $E_{\rm k}$ $\gg$ $E_{\rm total}$ and $E_{\rm total}$ $\ll$ $E_{0}$, it is safe to neglect the dynamic evolution of each layer and the energy injection of the magnetar to the afterglow jet. According to \cite{sunhui2023arXiv}, the initial X-ray luminosity is $L_{\rm X}(0) \sim 2 \times 10^{48}$~erg s$^{-1}$ during the prompt emission phase, and is of the same magnitude as $\xi L_{\rm sd}(0)$ in our fitting. Our result suggests that the presence of a magnetar in GRB~230307A is self-consistent in the prompt emission phase and the following multi-wavelength observations.

\section{Conclusion And Discussion} \label{sec:discussion_and_conclusion}

GRB~230307A is the first GRB with a magnetar engine signature in the prompt emission phase, the second brightest GRB ever observed (only dwarfed by GRB~221009A), and the third long GRB in association with suspected kilonova emission. The fitting results presented in this {\it Letter} demonstrate that the NS-WD merger with a magnetar remnant could be a self-consistent explanation for GRB~230307A both in the prompt emission and the following multi-wavelength afterglow including the optical excesses. The identification of NS-WD merger events from optical transient phenomena usually faces a challenge as their early electromagnetic radiation cannot be recorded initially, making it difficult to identify the nature of the remnant objects. By contrast, GRB 230307A offers a unique opportunity as it was observed by multiple satellites across various wavelengths and, in particular, the early prompt emission of the GRB provides the magnetar parameters, which enables us to test the consistency of the GRB~230307A observations with the NS-WD merger scenario.

By comparing with some simulation studies in literature \cite[e.g.,][]{Zenati2020MNRAS,Kaltenborn2022arXiv220913061K}, the NS-WD merger origin of GRB~230307A could further be supported by its detailed emission features. For example, \cite{Kaltenborn2022arXiv220913061K} showed a prominent emission line at 20000 \AA \ at 30.37~d, which can be found in the observation of GRB~230307A at 29~d. In the framework of the NS-WD merger model, the notable emission features of GRB 230307A at $\sim$ 2.1 $\mu$m (at 29 and 61 days) and $\sim$ 4.4 $\mu$m (at 61 days) could be attributed to the \ion{Ca}{2} line as well as the first overtone and second overtone features of the CO molecular line \citep{Levan2023, Gillanders2023arXiv}, although these features were previously ascribed to the $r$-process elements in \cite{Gillanders2020MNRAS}. Anyway, it should be noticed that our knowledge of the spectra of NS-WD mergers is actually very limited. For example, the early optical spectroscopy from 5000 \AA \ to 9200 \AA \ recorded by the Gemini South telescope demonstrated the absence of \ion{Si}{2} absorption lines around 6300 \AA \ at 2.4~d, which is however expected at early stage in NS-WD mergers \citep{Zenati2020MNRAS}. Such a situation also appeared in another NS-WD merger candidate event, AT 2018kzr, which showed no absorption lines above 5000 \AA \ at 1.9~d and a weak \ion{Si}{2} absorption line at 3.8~d \citep{Gillanders2020MNRAS}. In order to clarify these detailed emission features, a more elaborated model definitely needs to be considered in future works, by taking into account the complex nuclear heating and cooling, and the multigroup opacity in different energy band even in different region of the ejecta. Furthermore, the potential significant influence of the remnant magnetar on all of these physical properties should also be investigated carefully. 

Nevertheless, for GRB~230307A, actually only the peak of the ejecta emission can be resolved from the observational data, the characteristic of which is generally determined by the basic properties of the ejecta and energy sources (cf., the Arnett's law for supernova light curves \citep{Arnett1982ApJ}). Therefore, the simplified model can still be workable in our fittings. In future, it is expected that upcoming instruments like the Jonit Science Center for China Space Station Telescope (CSST) would be capable of measuring the spectra more precisely and, especially, the establishment of a collaborative network involving multiple telescopes would hold the potential to reveal more detailed features of the light curves and spectra of such kilonova-like transients. In particular, if the GRB afterglow emission can decay more quickly, then the ejecta emission would be detected at time much later than its peak, where the emission spectra could significantly deviate from the black body and many line features could appear. In that cases, the comparison between the observational data and a more elaborated model would lead to a more reliable constraint on the details of the nucleosynthesis and opacity of the ejecta, as investigated in theory in \cite{Kaltenborn2022arXiv220913061K} and \cite{Zenati2019MNRAS,Zenati2020MNRAS}.

Finally, we would like to point out that the formation of a system including a NS and a $^{56}$Ni-rich ejecta could also be caused by the accretion-induced collapse (AIC) of a WD in companion with a non-degenerate donor or another WD \citep{Nomoto1991ApJ, Taam1986ApJ, Duncan1992ApJ, Usov1992Natur, Yoon2007MNRAS}, which therefore could also lead to a bright optical transient \citep{Yu&Li2015ApJ, Yu&chen&wang2019ApJ, Yu&chen&li2019ApJ}. Nevertheless, the AIC processes are deemed less likely to produce GRBs; otherwise, the observed rate of GRBs would be too high to be consistent with the observations. This may hint that the magnetic fields of the AIC-formed NSs are relatively lower than those of the NS-WD merger products, according to the statistical inferences given by \cite{Yu&Zhu2017ApJ}. Besides the association with the GRB emission, the NS-WD mergers can also be different from the AIC events by their gravitational wave (GW) radiation, which would be detected by future millihertz and decihertz space-borne GW detectors \citep{Yinyihan2023ApJ, kangyacheng2023arXiv}.  In any case, it is still necessary to investigate the theoretical differences of the optical transients between the AIC events and NS-WD mergers. 

\section*{acknowledgements}
We thank Bin-Bin Zhang for reminding us to pay attention to this GRB, Yu-Han Yang for providing the data, Ken Chen and Jian-He Zheng for helpful discussions on the paper. This work is supported by the National Key Research and Development Programs of China (2021YFA0718500, 2022YFF0711404), the National SKA Program of China (2020SKA0120300, 2022SKA0130100, 2022SKA0130102), and the science research grants from the China Manned Space Project with NOs. CMS-CSST-2021-A12 and CMS-CSST-2021-B11.
This work was performed on an HPC server equipped with two Intel Xeon Gold 6248 modules at Nanjing University. We acknowledge IT support from the computer lab of the School of Astronomy and Space Science at Nanjing University.

%\bibliography{ms.bib}

\begin{thebibliography}{}
\expandafter\ifx\csname natexlab\endcsname\relax\def\natexlab#1{#1}\fi
\providecommand{\url}[1]{\href{#1}{#1}}
\providecommand{\dodoi}[1]{doi:~\href{http://doi.org/#1}{\nolinkurl{#1}}}
\providecommand{\doeprint}[1]{\href{http://ascl.net/#1}{\nolinkurl{http://ascl.net/#1}}}
\providecommand{\doarXiv}[1]{\href{https://arxiv.org/abs/#1}{\nolinkurl{https://arxiv.org/abs/#1}}}

\bibitem[{{Abbott} {et~al.}(2017{\natexlab{a}}){Abbott}, {Abbott}, {Abbott}, {Acernese}, {Ackley}, {Adams}, {Adams}, {Addesso}, {Adhikari}, {Adya}, {Affeldt}, {Afrough}, {Agarwal}, {Agathos}, {Agatsuma}, {Aggarwal}, {Aguiar}, {Aiello}, {Ain}, {Ajith}, {Allen}, {Allen}, {Allocca}, {Altin}, {Amato}, {Ananyeva}, {Anderson}, {Anderson}, {Angelova}, {Antier}, {Appert}, {Arai}, {Araya}, {Areeda}, {Arnaud}, {Arun}, {Ascenzi}, {Ashton}, {Ast}, {Aston}, {Astone}, {Atallah}, {Aufmuth}, {Aulbert}, {AultONeal}, {Austin}, {Avila-Alvarez}, {Babak}, {Bacon}, {Bader}, {Bae}, {Bailes}, {Baker}, {Baldaccini}, {Ballardin}, {Ballmer}, {Banagiri}, {Barayoga}, {Barclay}, {Barish}, {Barker}, {Barkett}, {Barone}, {Barr}, {Barsotti}, {Barsuglia}, {Barta}, {Barthelmy}, {Bartlett}, {Bartos}, {Bassiri}, {Basti}, {Batch}, {Bawaj}, {Bayley}, {Bazzan}, {B{\'e}csy}, {Beer}, {Bejger}, {Belahcene}, {Bell}, {Berger}, {Bergmann}, {Bernuzzi}, {Bero}, {Berry}, {Bersanetti}, {Bertolini}, {Betzwieser}, {Bhagwat}, {Bhandare}, {Bilenko},
  {Billingsley}, {Billman}, {Birch}, {Birney}, {Birnholtz}, {Biscans}, {Biscoveanu}, {Bisht}, {Bitossi}, {Biwer}, {Bizouard}, {Blackburn}, {Blackman}, {Blair}, {Blair}, {Blair}, {Bloemen}, {Bock}, {Bode}, {Boer}, {Bogaert}, {Bohe}, {Bondu}, {Bonilla}, {Bonnand}, {Boom}, {Bork}, {Boschi}, {Bose}, {Bossie}, {Bouffanais}, {Bozzi}, {Bradaschia}, {Brady}, {Branchesi}, {Brau}, {Briant}, {Brillet}, {Brinkmann}, {Brisson}, {Brockill}, {Broida}, {Brooks}, {Brown}, {Brown}, {Brunett}, {Buchanan}, {Buikema}, {Bulik}, {Bulten}, {Buonanno}, {Buskulic}, {Buy}, {Byer}, {Cabero}, {Cadonati}, {Cagnoli}, {Cahillane}, {Calder{\'o}n Bustillo}, {Callister}, {Calloni}, {Camp}, {Canepa}, {Canizares}, {Cannon}, {Cao}, {Cao}, {Capano}, {Capocasa}, {Carbognani}, {Caride}, {Carney}, {Carullo}, {Casanueva Diaz}, {Casentini}, {Caudill}, {Cavagli{\`a}}, {Cavalier}, {Cavalieri}, {Cella}, {Cepeda}, {Cerd{\'a}-Dur{\'a}n}, {Cerretani}, {Cesarini}, {Chamberlin}, {Chan}, {Chao}, {Charlton}, {Chase}, {Chassande-Mottin}, {Chatterjee},
  {Chatziioannou}, {Cheeseboro}, {Chen}, {Chen}, {Chen}, {Cheng}, {Chia}, {Chincarini}, {Chiummo}, {Chmiel}, {Cho}, {Cho}, {Chow}, {Christensen}, {Chu}, {Chua}, {Chua}, {Chung}, {Chung}, {Ciani}, {Ciolfi}, {Cirelli}, {Cirone}, {Clara}, {Clark}, {Clearwater}, {Cleva}, {Cocchieri}, {Coccia}, {Cohadon}, {Cohen}, {Colla}, {Collette}, {Cominsky}, {Constancio}, {Conti}, {Cooper}, {Corban}, {Corbitt}, {Cordero-Carri{\'o}n}, {Corley}, {Cornish}, {Corsi}, {Cortese}, {Costa}, {Coughlin}, {Coughlin}, {Coulon}, {Countryman}, {Couvares}, {Covas}, {Cowan}, {Coward}, {Cowart}, {Coyne}, {Coyne}, {Creighton}, {Creighton}, {Cripe}, {Crowder}, {Cullen}, {Cumming}, {Cunningham}, {Cuoco}, {Dal Canton}, {D{\'a}lya}, {Danilishin}, {D'Antonio}, {Danzmann}, {Dasgupta}, {Da Silva Costa}, {Dattilo}, {Dave}, {Davier}, {Davis}, {Daw}, {Day}, {De}, {DeBra}, {Degallaix}, {De Laurentis}, {Del{\'e}glise}, {Del Pozzo}, {Demos}, {Denker}, {Dent}, {De Pietri}, {Dergachev}, {De Rosa}, {DeRosa}, {De Rossi}, {DeSalvo}, {de Varona}, {Devenson},
  {Dhurandhar}, {D{\'\i}az}, {Dietrich}, {Di Fiore}, {Di Giovanni}, {Di Girolamo}, {Di Lieto}, {Di Pace}, {Di Palma}, {Di Renzo}, {Doctor}, {Dolique}, {Donovan}, {Dooley}, {Doravari}, {Dorrington}, {Douglas}, {Dovale {\'A}lvarez}, {Downes}, {Drago}, {Dreissigacker}, {Driggers}, {Du}, {Ducrot}, {Dudi}, {Dupej}, {Dwyer}, {Edo}, {Edwards}, {Effler}, {Eggenstein}, {Ehrens}, {Eichholz}, {Eikenberry}, {Eisenstein}, {Essick}, {Estevez}, {Etienne}, {Etzel}, {Evans}, {Evans}, {Factourovich}, {Fafone}, {Fair}, {Fairhurst}, {Fan}, {Farinon}, {Farr}, {Farr}, {Fauchon-Jones}, {Favata}, {Fays}, {Fee}, {Fehrmann}, {Feicht}, {Fejer}, {Fernandez-Galiana}, {Ferrante}, {Ferreira}, {Ferrini}, {Fidecaro}, {Finstad}, {Fiori}, {Fiorucci}, {Fishbach}, {Fisher}, {Fitz-Axen}, {Flaminio}, {Fletcher}, {Fong}, {Font}, {Forsyth}, {Forsyth}, {Fournier}, {Frasca}, {Frasconi}, {Frei}, {Freise}, {Frey}, {Frey}, {Fries}, {Fritschel}, {Frolov}, {Fulda}, {Fyffe}, {Gabbard}, {Gadre}, {Gaebel}, {Gair}, {Gammaitoni}, {Ganija}, {Gaonkar},
  {Garcia-Quiros}, {Garufi}, {Gateley}, {Gaudio}, {Gaur}, {Gayathri}, {Gehrels}, {Gemme}, {Genin}, {Gennai}, {George}, {George}, {Gergely}, {Germain}, {Ghonge}, {Ghosh}, {Ghosh}, {Ghosh}, {Giaime}, {Giardina}, {Giazotto}, {Gill}, {Glover}, {Goetz}, {Goetz}, {Gomes}, {Goncharov}, {Gonz{\'a}lez}, {Gonzalez Castro}, {Gopakumar}, {Gorodetsky}, {Gossan}, {Gosselin}, {Gouaty}, {Grado}, {Graef}, {Granata}, {Grant}, {Gras}, {Gray}, {Greco}, {Green}, {Gretarsson}, {Groot}, {Grote}, {Grunewald}, {Gruning}, {Guidi}, {Guo}, {Gupta}, {Gupta}, {Gushwa}, {Gustafson}, {Gustafson}, {Halim}, {Hall}, {Hall}, {Hamilton}, {Hammond}, {Haney}, {Hanke}, {Hanks}, {Hanna}, {Hannam}, {Hannuksela}, {Hanson}, {Hardwick}, {Harms}, {Harry}, {Harry}, {Hart}, {Haster}, {Haughian}, {Healy}, {Heidmann}, {Heintze}, {Heitmann}, {Hello}, {Hemming}, {Hendry}, {Heng}, {Hennig}, {Heptonstall}, {Heurs}, {Hild}, {Hinderer}, {Ho}, {Hoak}, {Hofman}, {Holt}, {Holz}, {Hopkins}, {Horst}, {Hough}, {Houston}, {Howell}, {Hreibi}, {Hu}, {Huerta}, {Huet},
  {Hughey}, {Husa}, {Huttner}, {Huynh-Dinh}, {Indik}, {Inta}, {Intini}, {Isa}, {Isac}, {Isi}, {Iyer}, {Izumi}, {Jacqmin}, {Jani}, {Jaranowski}, {Jawahar}, {Jim{\'e}nez-Forteza}, {Johnson}, {Johnson-McDaniel}, {Jones}, {Jones}, {Jonker}, {Ju}, {Junker}, {Kalaghatgi}, {Kalogera}, {Kamai}, {Kandhasamy}, {Kang}, {Kanner}, {Kapadia}, {Karki}, {Karvinen}, {Kasprzack}, {Kastaun}, {Katolik}, {Katsavounidis}, {Katzman}, {Kaufer}, {Kawabe}, {K{\'e}f{\'e}lian}, {Keitel}, {Kemball}, {Kennedy}, {Kent}, {Key}, {Khalili}, {Khan}, {Khan}, {Khan}, {Khazanov}, {Kijbunchoo}, {Kim}, {Kim}, {Kim}, {Kim}, {Kim}, {Kim}, {Kimbrell}, {King}, {King}, {Kinley-Hanlon}, {Kirchhoff}, {Kissel}, {Kleybolte}, {Klimenko}, {Knowles}, {Koch}, {Koehlenbeck}, {Koley}, {Kondrashov}, {Kontos}, {Korobko}, {Korth}, {Kowalska}, {Kozak}, {Kr{\"a}mer}, {Kringel}, {Krishnan}, {Kr{\'o}lak}, {Kuehn}, {Kumar}, {Kumar}, {Kumar}, {Kuo}, {Kutynia}, {Kwang}, {Lackey}, {Lai}, {Landry}, {Lang}, {Lange}, {Lantz}, {Lanza}, {Larson}, {Lartaux-Vollard}, {Lasky},
  {Laxen}, {Lazzarini}, {Lazzaro}, {Leaci}, {Leavey}, {Lee}, {Lee}, {Lee}, {Lee}, {Lee}, {Lehmann}, {Lenon}, {Leon}, {Leonardi}, {Leroy}, {Letendre}, {Levin}, {Li}, {Linker}, {Littenberg}, {Liu}, {Liu}, {Lo}, {Lockerbie}, {London}, {Lord}, {Lorenzini}, {Loriette}, {Lormand}, {Losurdo}, {Lough}, {Lousto}, {Lovelace}, {L{\"u}ck}, {Lumaca}, {Lundgren}, {Lynch}, {Ma}, {Macas}, {Macfoy}, {Machenschalk}, {MacInnis}, {Macleod}, {Maga{\~n}a Hernandez}, {Maga{\~n}a-Sandoval}, {Maga{\~n}a Zertuche}, {Magee}, {Majorana}, {Maksimovic}, {Man}, {Mandic}, {Mangano}, {Mansell}, {Manske}, {Mantovani}, {Marchesoni}, {Marion}, {M{\'a}rka}, {M{\'a}rka}, {Markakis}, {Markosyan}, {Markowitz}, {Maros}, {Marquina}, {Marsh}, {Martelli}, {Martellini}, {Martin}, {Martin}, {Martynov}, {Marx}, {Mason}, {Massera}, {Masserot}, {Massinger}, {Masso-Reid}, {Mastrogiovanni}, {Matas}, {Matichard}, {Matone}, {Mavalvala}, {Mazumder}, {McCarthy}, {McClelland}, {McCormick}, {McCuller}, {McGuire}, {McIntyre}, {McIver}, {McManus}, {McNeill}, {McRae},
  {McWilliams}, {Meacher}, {Meadors}, {Mehmet}, {Meidam}, {Mejuto-Villa}, {Melatos}, {Mendell}, {Mercer}, {Merilh}, {Merzougui}, {Meshkov}, {Messenger}, {Messick}, {Metzdorff}, {Meyers}, {Miao}, {Michel}, {Middleton}, {Mikhailov}, {Milano}, {Miller}, {Miller}, {Miller}, {Millhouse}, {Milovich-Goff}, {Minazzoli}, {Minenkov}, {Ming}, {Mishra}, {Mitra}, {Mitrofanov}, {Mitselmakher}, {Mittleman}, {Moffa}, {Moggi}, {Mogushi}, {Mohan}, {Mohapatra}, {Molina}, {Montani}, {Moore}, {Moraru}, {Moreno}, {Morisaki}, {Morriss}, {Mours}, {Mow-Lowry}, {Mueller}, {Muir}, {Mukherjee}, {Mukherjee}, {Mukherjee}, {Mukund}, {Mullavey}, {Munch}, {Mu{\~n}iz}, {Muratore}, {Murray}, {Nagar}, {Napier}, {Nardecchia}, {Naticchioni}, {Nayak}, {Neilson}, {Nelemans}, {Nelson}, {Nery}, {Neunzert}, {Nevin}, {Newport}, {Newton}, {Ng}, {Nguyen}, {Nguyen}, {Nichols}, {Nielsen}, {Nissanke}, {Nitz}, {Noack}, {Nocera}, {Nolting}, {North}, {Nuttall}, {Oberling}, {O'Dea}, {Ogin}, {Oh}, {Oh}, {Ohme}, {Okada}, {Oliver}, {Oppermann}, {Oram}, {O'Reilly},
  {Ormiston}, {Ortega}, {O'Shaughnessy}, {Ossokine}, {Ottaway}, {Overmier}, {Owen}, {Pace}, {Page}, {Page}, {Pai}, {Pai}, {Palamos}, {Palashov}, {Palomba}, {Pal-Singh}, {Pan}, {Pan}, {Pang}, {Pang}, {Pankow}, {Pannarale}, {Pant}, {Paoletti}, {Paoli}, {Papa}, {Parida}, {Parker}, {Pascucci}, {Pasqualetti}, {Passaquieti}, {Passuello}, {Patil}, {Patricelli}, {Pearlstone}, {Pedraza}, {Pedurand}, {Pekowsky}, {Pele}, {Penn}, {Perez}, {Perreca}, {Perri}, {Pfeiffer}, {Phelps}, {Piccinni}, {Pichot}, {Piergiovanni}, {Pierro}, {Pillant}, {Pinard}, {Pinto}, {Pirello}, {Pitkin}, {Poe}, {Poggiani}, {Popolizio}, {Porter}, {Post}, {Powell}, {Prasad}, {Pratt}, {Pratten}, {Predoi}, {Prestegard}, {Prijatelj}, {Principe}, {Privitera}, {Prix}, {Prodi}, {Prokhorov}, {Puncken}, {Punturo}, {Puppo}, {P{\"u}rrer}, {Qi}, {Quetschke}, {Quintero}, {Quitzow-James}, {Raab}, {Rabeling}, {Radkins}, {Raffai}, {Raja}, {Rajan}, {Rajbhandari}, {Rakhmanov}, {Ramirez}, {Ramos-Buades}, {Rapagnani}, {Raymond}, {Razzano}, {Read}, {Regimbau}, {Rei},
  {Reid}, {Reitze}, {Ren}, {Reyes}, {Ricci}, {Ricker}, {Rieger}, {Riles}, {Rizzo}, {Robertson}, {Robie}, {Robinet}, {Rocchi}, {Rolland}, {Rollins}, {Roma}, {Romano}, {Romano}, {Romel}, {Romie}, {Rosi{\'n}ska}, {Ross}, {Rowan}, {R{\"u}diger}, {Ruggi}, {Rutins}, {Ryan}, {Sachdev}, {Sadecki}, {Sadeghian}, {Sakellariadou}, {Salconi}, {Saleem}, {Salemi}, {Samajdar}, {Sammut}, {Sampson}, {Sanchez}, {Sanchez}, {Sanchis-Gual}, {Sandberg}, {Sanders}, {Sassolas}, {Sathyaprakash}, {Saulson}, {Sauter}, {Savage}, {Sawadsky}, {Schale}, {Scheel}, {Scheuer}, {Schmidt}, {Schmidt}, {Schnabel}, {Schofield}, {Sch{\"o}nbeck}, {Schreiber}, {Schuette}, {Schulte}, {Schutz}, {Schwalbe}, {Scott}, {Scott}, {Seidel}, {Sellers}, {Sengupta}, {Sentenac}, {Sequino}, {Sergeev}, {Shaddock}, {Shaffer}, {Shah}, {Shahriar}, {Shaner}, {Shao}, {Shapiro}, {Shawhan}, {Sheperd}, {Shoemaker}, {Shoemaker}, {Siellez}, {Siemens}, {Sieniawska}, {Sigg}, {Silva}, {Singer}, {Singh}, {Singhal}, {Sintes}, {Slagmolen}, {Smith}, {Smith}, {Smith}, {Somala},
  {Son}, {Sonnenberg}, {Sorazu}, {Sorrentino}, {Souradeep}, {Spencer}, {Srivastava}, {Staats}, {Staley}, {Steinke}, {Steinlechner}, {Steinlechner}, {Steinmeyer}, {Stevenson}, {Stone}, {Stops}, {Strain}, {Stratta}, {Strigin}, {Strunk}, {Sturani}, {Stuver}, {Summerscales}, {Sun}, {Sunil}, {Suresh}, {Sutton}, {Swinkels}, {Szczepa{\'n}czyk}, {Tacca}, {Tait}, {Talbot}, {Talukder}, {Tanner}, {T{\'a}pai}, {Taracchini}, {Tasson}, {Taylor}, {Taylor}, {Tewari}, {Theeg}, {Thies}, {Thomas}, {Thomas}, {Thomas}, {Thorne}, {Thorne}, {Thrane}, {Tiwari}, {Tiwari}, {Tokmakov}, {Toland}, {Tonelli}, {Tornasi}, {Torres-Forn{\'e}}, {Torrie}, {T{\"o}yr{\"a}}, {Travasso}, {Traylor}, {Trinastic}, {Tringali}, {Trozzo}, {Tsang}, {Tse}, {Tso}, {Tsukada}, {Tsuna}, {Tuyenbayev}, {Ueno}, {Ugolini}, {Unnikrishnan}, {Urban}, {Usman}, {Vahlbruch}, {Vajente}, {Valdes}, {Vallisneri}, {van Bakel}, {van Beuzekom}, {van den Brand}, {Van Den Broeck}, {Vander-Hyde}, {van der Schaaf}, {van Heijningen}, {van Veggel}, {Vardaro}, {Varma}, {Vass},
  {Vas{\'u}th}, {Vecchio}, {Vedovato}, {Veitch}, {Veitch}, {Venkateswara}, {Venugopalan}, {Verkindt}, {Vetrano}, {Vicer{\'e}}, {Viets}, {Vinciguerra}, {Vine}, {Vinet}, {Vitale}, {Vo}, {Vocca}, {Vorvick}, {Vyatchanin}, {Wade}, {Wade}, {Wade}, {Walet}, {Walker}, {Wallace}, {Walsh}, {Wang}, {Wang}, {Wang}, {Wang}, {Wang}, {Ward}, {Warner}, {Was}, {Watchi}, {Weaver}, {Wei}, {Weinert}, {Weinstein}, {Weiss}, {Wen}, {Wessel}, {We{\ss}els}, {Westerweck}, {Westphal}, {Wette}, {Whelan}, {Whitcomb}, {Whiting}, {Whittle}, {Wilken}, {Williams}, {Williams}, {Williamson}, {Willis}, {Willke}, {Wimmer}, {Winkler}, {Wipf}, {Wittel}, {Woan}, {Woehler}, {Wofford}, {Wong}, {Worden}, {Wright}, {Wu}, {Wysocki}, {Xiao}, {Yamamoto}, {Yancey}, {Yang}, {Yap}, {Yazback}, {Yu}, {Yu}, {Yvert}, {Zadro{\.Z}ny}, {Zanolin}, {Zelenova}, {Zendri}, {Zevin}, {Zhang}, {Zhang}, {Zhang}, {Zhang}, {Zhao}, {Zhou}, {Zhou}, {Zhu}, {Zhu}, {Zimmerman}, {Zucker}, {Zweizig}, {LIGO Scientific Collaboration}, \& {Virgo Collaboration}}]{Abbott2017PhRvL}
{Abbott}, B.~P., {Abbott}, R., {Abbott}, T.~D., {et~al.} 2017{\natexlab{a}}, \prl, 119, 161101, \dodoi{10.1103/PhysRevLett.119.161101}

\bibitem[{{Abbott} {et~al.}(2017{\natexlab{b}}){Abbott}, {Abbott}, {Abbott}, {Acernese}, {Ackley}, {Adams}, {Adams}, {Addesso}, {Adhikari}, {Adya}, {Affeldt}, {Afrough}, {Agarwal}, {Agathos}, {Agatsuma}, {Aggarwal}, {Aguiar}, {Aiello}, {Ain}, {Ajith}, {Allen}, {Allen}, {Allocca}, {Aloy}, {Altin}, {Amato}, {Ananyeva}, {Anderson}, {Anderson}, {Angelova}, {Antier}, {Appert}, {Arai}, {Araya}, {Areeda}, {Arnaud}, {Arun}, {Ascenzi}, {Ashton}, {Ast}, {Aston}, {Astone}, {Atallah}, {Aufmuth}, {Aulbert}, {AultONeal}, {Austin}, {Avila-Alvarez}, {Babak}, {Bacon}, {Bader}, {Bae}, {Baker}, {Baldaccini}, {Ballardin}, {Ballmer}, {Banagiri}, {Barayoga}, {Barclay}, {Barish}, {Barker}, {Barkett}, {Barone}, {Barr}, {Barsotti}, {Barsuglia}, {Barta}, {Bartlett}, {Bartos}, {Bassiri}, {Basti}, {Batch}, {Bawaj}, {Bayley}, {Bazzan}, {B{\'e}csy}, {Beer}, {Bejger}, {Belahcene}, {Bell}, {Berger}, {Bergmann}, {Bero}, {Berry}, {Bersanetti}, {Bertolini}, {Betzwieser}, {Bhagwat}, {Bhandare}, {Bilenko}, {Billingsley}, {Billman}, {Birch},
  {Birney}, {Birnholtz}, {Biscans}, {Biscoveanu}, {Bisht}, {Bitossi}, {Biwer}, {Bizouard}, {Blackburn}, {Blackman}, {Blair}, {Blair}, {Blair}, {Bloemen}, {Bock}, {Bode}, {Boer}, {Bogaert}, {Bohe}, {Bondu}, {Bonilla}, {Bonnand}, {Boom}, {Bork}, {Boschi}, {Bose}, {Bossie}, {Bouffanais}, {Bozzi}, {Bradaschia}, {Brady}, {Branchesi}, {Brau}, {Briant}, {Brillet}, {Brinkmann}, {Brisson}, {Brockill}, {Broida}, {Brooks}, {Brown}, {Brown}, {Brunett}, {Buchanan}, {Buikema}, {Bulik}, {Bulten}, {Buonanno}, {Buskulic}, {Buy}, {Byer}, {Cabero}, {Cadonati}, {Cagnoli}, {Cahillane}, {Calder{\'o}n Bustillo}, {Callister}, {Calloni}, {Camp}, {Canepa}, {Canizares}, {Cannon}, {Cao}, {Cao}, {Capano}, {Capocasa}, {Carbognani}, {Caride}, {Carney}, {Casanueva Diaz}, {Casentini}, {Caudill}, {Cavagli{\`a}}, {Cavalier}, {Cavalieri}, {Cella}, {Cepeda}, {Cerd{\'a}-Dur{\'a}n}, {Cerretani}, {Cesarini}, {Chamberlin}, {Chan}, {Chao}, {Charlton}, {Chase}, {Chassande-Mottin}, {Chatterjee}, {Chatziioannou}, {Cheeseboro}, {Chen}, {Chen}, {Chen},
  {Cheng}, {Chia}, {Chincarini}, {Chiummo}, {Chmiel}, {Cho}, {Cho}, {Chow}, {Christensen}, {Chu}, {Chua}, {Chua}, {Chung}, {Chung}, {Ciani}, {Ciolfi}, {Cirelli}, {Cirone}, {Clara}, {Clark}, {Clearwater}, {Cleva}, {Cocchieri}, {Coccia}, {Cohadon}, {Cohen}, {Colla}, {Collette}, {Cominsky}, {Constancio}, {Conti}, {Cooper}, {Corban}, {Corbitt}, {Cordero-Carri{\'o}n}, {Corley}, {Cornish}, {Corsi}, {Cortese}, {Costa}, {Coughlin}, {Coughlin}, {Coulon}, {Countryman}, {Couvares}, {Covas}, {Cowan}, {Coward}, {Cowart}, {Coyne}, {Coyne}, {Creighton}, {Creighton}, {Cripe}, {Crowder}, {Cullen}, {Cumming}, {Cunningham}, {Cuoco}, {Dal Canton}, {D{\'a}lya}, {Danilishin}, {D'Antonio}, {Danzmann}, {Dasgupta}, {Da Silva Costa}, {Dattilo}, {Dave}, {Davier}, {Davis}, {Daw}, {Day}, {De}, {DeBra}, {Degallaix}, {De Laurentis}, {Del{\'e}glise}, {Del Pozzo}, {Demos}, {Denker}, {Dent}, {De Pietri}, {Dergachev}, {De Rosa}, {DeRosa}, {De Rossi}, {DeSalvo}, {de Varona}, {Devenson}, {Dhurandhar}, {D{\'\i}az}, {Di Fiore}, {Di Giovanni}, {Di
  Girolamo}, {Di Lieto}, {Di Pace}, {Di Palma}, {Di Renzo}, {Doctor}, {Dolique}, {Donovan}, {Dooley}, {Doravari}, {Dorrington}, {Douglas}, {Dovale {\'A}lvarez}, {Downes}, {Drago}, {Dreissigacker}, {Driggers}, {Du}, {Ducrot}, {Dupej}, {Dwyer}, {Edo}, {Edwards}, {Effler}, {Eggenstein}, {Ehrens}, {Eichholz}, {Eikenberry}, {Eisenstein}, {Essick}, {Estevez}, {Etienne}, {Etzel}, {Evans}, {Evans}, {Factourovich}, {Fafone}, {Fair}, {Fairhurst}, {Fan}, {Farinon}, {Farr}, {Farr}, {Fauchon-Jones}, {Favata}, {Fays}, {Fee}, {Fehrmann}, {Feicht}, {Fejer}, {Fernandez-Galiana}, {Ferrante}, {Ferreira}, {Ferrini}, {Fidecaro}, {Finstad}, {Fiori}, {Fiorucci}, {Fishbach}, {Fisher}, {Fitz-Axen}, {Flaminio}, {Fletcher}, {Fong}, {Font}, {Forsyth}, {Forsyth}, {Fournier}, {Frasca}, {Frasconi}, {Frei}, {Freise}, {Frey}, {Frey}, {Fries}, {Fritschel}, {Frolov}, {Fulda}, {Fyffe}, {Gabbard}, {Gadre}, {Gaebel}, {Gair}, {Gammaitoni}, {Ganija}, {Gaonkar}, {Garcia-Quiros}, {Garufi}, {Gateley}, {Gaudio}, {Gaur}, {Gayathri}, {Gehrels}, {Gemme},
  {Genin}, {Gennai}, {George}, {George}, {Gergely}, {Germain}, {Ghonge}, {Ghosh}, {Ghosh}, {Ghosh}, {Giaime}, {Giardina}, {Giazotto}, {Gill}, {Glover}, {Goetz}, {Goetz}, {Gomes}, {Goncharov}, {Gonz{\'a}lez}, {Gonzalez Castro}, {Gopakumar}, {Gorodetsky}, {Gossan}, {Gosselin}, {Gouaty}, {Grado}, {Graef}, {Granata}, {Grant}, {Gras}, {Gray}, {Greco}, {Green}, {Gretarsson}, {Groot}, {Grote}, {Grunewald}, {Gruning}, {Guidi}, {Guo}, {Gupta}, {Gupta}, {Gushwa}, {Gustafson}, {Gustafson}, {Halim}, {Hall}, {Hall}, {Hamilton}, {Hammond}, {Haney}, {Hanke}, {Hanks}, {Hanna}, {Hannam}, {Hannuksela}, {Hanson}, {Hardwick}, {Harms}, {Harry}, {Harry}, {Hart}, {Haster}, {Haughian}, {Healy}, {Heidmann}, {Heintze}, {Heitmann}, {Hello}, {Hemming}, {Hendry}, {Heng}, {Hennig}, {Heptonstall}, {Heurs}, {Hild}, {Hinderer}, {Hoak}, {Hofman}, {Holt}, {Holz}, {Hopkins}, {Horst}, {Hough}, {Houston}, {Howell}, {Hreibi}, {Hu}, {Huerta}, {Huet}, {Hughey}, {Husa}, {Huttner}, {Huynh-Dinh}, {Indik}, {Inta}, {Intini}, {Isa}, {Isac}, {Isi}, {Iyer},
  {Izumi}, {Jacqmin}, {Jani}, {Jaranowski}, {Jawahar}, {Jim{\'e}nez-Forteza}, {Johnson}, {Johnson-McDaniel}, {Jones}, {Jones}, {Jonker}, {Ju}, {Junker}, {Kalaghatgi}, {Kalogera}, {Kamai}, {Kandhasamy}, {Kang}, {Kanner}, {Kapadia}, {Karki}, {Karvinen}, {Kasprzack}, {Kastaun}, {Katolik}, {Katsavounidis}, {Katzman}, {Kaufer}, {Kawabe}, {K{\'e}f{\'e}lian}, {Keitel}, {Kemball}, {Kennedy}, {Kent}, {Key}, {Khalili}, {Khan}, {Khan}, {Khan}, {Khazanov}, {Kijbunchoo}, {Kim}, {Kim}, {Kim}, {Kim}, {Kim}, {Kim}, {Kimbrell}, {King}, {King}, {Kinley-Hanlon}, {Kirchhoff}, {Kissel}, {Kleybolte}, {Klimenko}, {Knowles}, {Koch}, {Koehlenbeck}, {Koley}, {Kondrashov}, {Kontos}, {Korobko}, {Korth}, {Kowalska}, {Kozak}, {Kr{\"a}mer}, {Kringel}, {Krishnan}, {Kr{\'o}lak}, {Kuehn}, {Kumar}, {Kumar}, {Kumar}, {Kuo}, {Kutynia}, {Kwang}, {Lackey}, {Lai}, {Landry}, {Lang}, {Lange}, {Lantz}, {Lanza}, {Lartaux-Vollard}, {Lasky}, {Laxen}, {Lazzarini}, {Lazzaro}, {Leaci}, {Leavey}, {Lee}, {Lee}, {Lee}, {Lee}, {Lee}, {Lehmann}, {Lenon},
  {Leonardi}, {Leroy}, {Letendre}, {Levin}, {Li}, {Linker}, {Littenberg}, {Liu}, {Lo}, {Lockerbie}, {London}, {Lord}, {Lorenzini}, {Loriette}, {Lormand}, {Losurdo}, {Lough}, {Lousto}, {Lovelace}, {L{\"u}ck}, {Lumaca}, {Lundgren}, {Lynch}, {Ma}, {Macas}, {Macfoy}, {Machenschalk}, {MacInnis}, {Macleod}, {Maga{\~n}a Hernandez}, {Maga{\~n}a-Sandoval}, {Maga{\~n}a Zertuche}, {Magee}, {Majorana}, {Maksimovic}, {Man}, {Mandic}, {Mangano}, {Mansell}, {Manske}, {Mantovani}, {Marchesoni}, {Marion}, {M{\'a}rka}, {M{\'a}rka}, {Markakis}, {Markosyan}, {Markowitz}, {Maros}, {Marquina}, {Martelli}, {Martellini}, {Martin}, {Martin}, {Martynov}, {Mason}, {Massera}, {Masserot}, {Massinger}, {Masso-Reid}, {Mastrogiovanni}, {Matas}, {Matichard}, {Matone}, {Mavalvala}, {Mazumder}, {McCarthy}, {McClelland}, {McCormick}, {McCuller}, {McGuire}, {McIntyre}, {McIver}, {McManus}, {McNeill}, {McRae}, {McWilliams}, {Meacher}, {Meadors}, {Mehmet}, {Meidam}, {Mejuto-Villa}, {Melatos}, {Mendell}, {Mercer}, {Merilh}, {Merzougui}, {Meshkov},
  {Messenger}, {Messick}, {Metzdorff}, {Meyers}, {Miao}, {Michel}, {Middleton}, {Mikhailov}, {Milano}, {Miller}, {Miller}, {Miller}, {Millhouse}, {Milovich-Goff}, {Minazzoli}, {Minenkov}, {Ming}, {Mishra}, {Mitra}, {Mitrofanov}, {Mitselmakher}, {Mittleman}, {Moffa}, {Moggi}, {Mogushi}, {Mohan}, {Mohapatra}, {Montani}, {Moore}, {Moraru}, {Moreno}, {Morriss}, {Mours}, {Mow-Lowry}, {Mueller}, {Muir}, {Mukherjee}, {Mukherjee}, {Mukherjee}, {Mukund}, {Mullavey}, {Munch}, {Mu{\~n}iz}, {Muratore}, {Murray}, {Napier}, {Nardecchia}, {Naticchioni}, {Nayak}, {Neilson}, {Nelemans}, {Nelson}, {Nery}, {Neunzert}, {Nevin}, {Newport}, {Newton}, {Ng}, {Nguyen}, {Nichols}, {Nielsen}, {Nissanke}, {Nitz}, {Noack}, {Nocera}, {Nolting}, {North}, {Nuttall}, {Oberling}, {O'Dea}, {Ogin}, {Oh}, {Oh}, {Ohme}, {Okada}, {Oliver}, {Oppermann}, {Oram}, {O'Reilly}, {Ormiston}, {Ortega}, {O'Shaughnessy}, {Ossokine}, {Ottaway}, {Overmier}, {Owen}, {Pace}, {Page}, {Page}, {Pai}, {Pai}, {Palamos}, {Palashov}, {Palomba}, {Pal-Singh}, {Pan},
  {Pan}, {Pang}, {Pang}, {Pankow}, {Pannarale}, {Pant}, {Paoletti}, {Paoli}, {Papa}, {Parida}, {Parker}, {Pascucci}, {Pasqualetti}, {Passaquieti}, {Passuello}, {Patil}, {Patricelli}, {Pearlstone}, {Pedraza}, {Pedurand}, {Pekowsky}, {Pele}, {Penn}, {Perez}, {Perreca}, {Perri}, {Pfeiffer}, {Phelps}, {Piccinni}, {Pichot}, {Piergiovanni}, {Pierro}, {Pillant}, {Pinard}, {Pinto}, {Pirello}, {Pitkin}, {Poe}, {Poggiani}, {Popolizio}, {Porter}, {Post}, {Powell}, {Prasad}, {Pratt}, {Pratten}, {Predoi}, {Prestegard}, {Prijatelj}, {Principe}, {Privitera}, {Prodi}, {Prokhorov}, {Puncken}, {Punturo}, {Puppo}, {P{\"u}rrer}, {Qi}, {Quetschke}, {Quintero}, {Quitzow-James}, {Raab}, {Rabeling}, {Radkins}, {Raffai}, {Raja}, {Rajan}, {Rajbhandari}, {Rakhmanov}, {Ramirez}, {Ramos-Buades}, {Rapagnani}, {Raymond}, {Razzano}, {Read}, {Regimbau}, {Rei}, {Reid}, {Reitze}, {Ren}, {Reyes}, {Ricci}, {Ricker}, {Rieger}, {Riles}, {Rizzo}, {Robertson}, {Robie}, {Robinet}, {Rocchi}, {Rolland}, {Rollins}, {Roma}, {Romano}, {Romel}, {Romie},
  {Rosi{\'n}ska}, {Ross}, {Rowan}, {R{\"u}diger}, {Ruggi}, {Rutins}, {Ryan}, {Sachdev}, {Sadecki}, {Sadeghian}, {Sakellariadou}, {Salconi}, {Saleem}, {Salemi}, {Samajdar}, {Sammut}, {Sampson}, {Sanchez}, {Sanchez}, {Sanchis-Gual}, {Sandberg}, {Sanders}, {Sassolas}, {Sathyaprakash}, {Saulson}, {Sauter}, {Savage}, {Sawadsky}, {Schale}, {Scheel}, {Scheuer}, {Schmidt}, {Schmidt}, {Schnabel}, {Schofield}, {Sch{\"o}nbeck}, {Schreiber}, {Schuette}, {Schulte}, {Schutz}, {Schwalbe}, {Scott}, {Scott}, {Seidel}, {Sellers}, {Sengupta}, {Sentenac}, {Sequino}, {Sergeev}, {Shaddock}, {Shaffer}, {Shah}, {Shahriar}, {Shaner}, {Shao}, {Shapiro}, {Shawhan}, {Sheperd}, {Shoemaker}, {Shoemaker}, {Siellez}, {Siemens}, {Sieniawska}, {Sigg}, {Silva}, {Singer}, {Singh}, {Singhal}, {Sintes}, {Slagmolen}, {Smith}, {Smith}, {Smith}, {Somala}, {Son}, {Sonnenberg}, {Sorazu}, {Sorrentino}, {Souradeep}, {Spencer}, {Srivastava}, {Staats}, {Staley}, {Steinke}, {Steinlechner}, {Steinlechner}, {Steinmeyer}, {Stevenson}, {Stone}, {Stops},
  {Strain}, {Stratta}, {Strigin}, {Strunk}, {Sturani}, {Stuver}, {Summerscales}, {Sun}, {Sunil}, {Suresh}, {Sutton}, {Swinkels}, {Szczepa{\'n}czyk}, {Tacca}, {Tait}, {Talbot}, {Talukder}, {Tanner}, {T{\'a}pai}, {Taracchini}, {Tasson}, {Taylor}, {Taylor}, {Tewari}, {Theeg}, {Thies}, {Thomas}, {Thomas}, {Thomas}, {Thorne}, {Thorne}, {Thrane}, {Tiwari}, {Tiwari}, {Tokmakov}, {Toland}, {Tonelli}, {Tornasi}, {Torres-Forn{\'e}}, {Torrie}, {T{\"o}yr{\"a}}, {Travasso}, {Traylor}, {Trinastic}, {Tringali}, {Trozzo}, {Tsang}, {Tse}, {Tso}, {Tsukada}, {Tsuna}, {Tuyenbayev}, {Ueno}, {Ugolini}, {Unnikrishnan}, {Urban}, {Usman}, {Vahlbruch}, {Vajente}, {Valdes}, {van Bakel}, {van Beuzekom}, {van den Brand}, {Van Den Broeck}, {Vander-Hyde}, {van der Schaaf}, {van Heijningen}, {van Veggel}, {Vardaro}, {Varma}, {Vass}, {Vas{\'u}th}, {Vecchio}, {Vedovato}, {Veitch}, {Veitch}, {Venkateswara}, {Venugopalan}, {Verkindt}, {Vetrano}, {Vicer{\'e}}, {Viets}, {Vinciguerra}, {Vine}, {Vinet}, {Vitale}, {Vo}, {Vocca}, {Vorvick},
  {Vyatchanin}, {Wade}, {Wade}, {Wade}, {Walet}, {Walker}, {Wallace}, {Walsh}, {Wang}, {Wang}, {Wang}, {Wang}, {Wang}, {Ward}, {Warner}, {Was}, {Watchi}, {Weaver}, {Wei}, {Weinert}, {Weinstein}, {Weiss}, {Wen}, {Wessel}, {We{\ss}els}, {Westerweck}, {Westphal}, {Wette}, {Whelan}, {Whitcomb}, {Whiting}, {Whittle}, {Wilken}, {Williams}, {Williams}, {Williamson}, {Willis}, {Willke}, {Wimmer}, {Winkler}, {Wipf}, {Wittel}, {Woan}, {Woehler}, {Wofford}, {Wong}, {Worden}, {Wright}, {Wu}, {Wysocki}, {Xiao}, {Yamamoto}, {Yancey}, {Yang}, {Yap}, {Yazback}, {Yu}, {Yu}, {Yvert}, {Zadro{\.z}ny}, {Zanolin}, {Zelenova}, {Zendri}, {Zevin}, {Zhang}, {Zhang}, {Zhang}, {Zhang}, {Zhao}, {Zhou}, {Zhou}, {Zhu}, {Zhu}, {Zimmerman}, {Zucker}, {Zweizig}, {(LIGO Scientific Collaboration}, {Virgo Collaboration}, {Burns}, {Veres}, {Kocevski}, {Racusin}, {Goldstein}, {Connaughton}, {Briggs}, {Blackburn}, {Hamburg}, {Hui}, {von Kienlin}, {McEnery}, {Preece}, {Wilson-Hodge}, {Bissaldi}, {Cleveland}, {Gibby}, {Giles}, {Kippen}, {McBreen},
  {Meegan}, {Paciesas}, {Poolakkil}, {Roberts}, {Stanbro}, {Gamma-ray Burst Monitor}, {Savchenko}, {Ferrigno}, {Kuulkers}, {Bazzano}, {Bozzo}, {Brandt}, {Chenevez}, {Courvoisier}, {Diehl}, {Domingo}, {Hanlon}, {Jourdain}, {Laurent}, {Lebrun}, {Lutovinov}, {Mereghetti}, {Natalucci}, {Rodi}, {Roques}, {Sunyaev}, {Ubertini}, \& {(INTEGRAL}}]{Abbott2017ApJ...848L..13A}
---. 2017{\natexlab{b}}, \apjl, 848, L13, \dodoi{10.3847/2041-8213/aa920c}

\bibitem[{{Ai} {et~al.}(2022){Ai}, {Zhang}, \& {Zhu}}]{ShunkeAi2022MNRAS}
{Ai}, S., {Zhang}, B., \& {Zhu}, Z. 2022, \mnras, 516, 2614, \dodoi{10.1093/mnras/stac2380}

\bibitem[{{Arcavi} {et~al.}(2017){Arcavi}, {Hosseinzadeh}, {Howell}, {McCully}, {Poznanski}, {Kasen}, {Barnes}, {Zaltzman}, {Vasylyev}, {Maoz}, \& {Valenti}}]{Arcavi2017Natur}
{Arcavi}, I., {Hosseinzadeh}, G., {Howell}, D.~A., {et~al.} 2017, \nat, 551, 64, \dodoi{10.1038/nature24291}

\bibitem[{{Arnett}(1982)}]{Arnett1982ApJ}
{Arnett}, W.~D. 1982, \apj, 253, 785, \dodoi{10.1086/159681}

\bibitem[{{Bartos} {et~al.}(2013){Bartos}, {Brady}, \& {M{\'a}rka}}]{Bartos2013CQGra}
{Bartos}, I., {Brady}, P., \& {M{\'a}rka}, S. 2013, Classical and Quantum Gravity, 30, 123001, \dodoi{10.1088/0264-9381/30/12/123001}

\bibitem[{{Blinnikov} {et~al.}(2006){Blinnikov}, {R{\"o}pke}, {Sorokina}, {Gieseler}, {Reinecke}, {Travaglio}, {Hillebrandt}, \& {Stritzinger}}]{Blinnikov2006A&A}
{Blinnikov}, S.~I., {R{\"o}pke}, F.~K., {Sorokina}, E.~I., {et~al.} 2006, \aap, 453, 229, \dodoi{10.1051/0004-6361:20054594}

\bibitem[{{Bloom} {et~al.}(1999){Bloom}, {Kulkarni}, {Djorgovski}, {Eichelberger}, {C{\^o}t{\'e}}, {Blakeslee}, {Odewahn}, {Harrison}, {Frail}, {Filippenko}, {Leonard}, {Riess}, {Spinrad}, {Stern}, {Bunker}, {Dey}, {Grossan}, {Perlmutter}, {Knop}, {Hook}, \& {Feroci}}]{Bloom1999Natur}
{Bloom}, J.~S., {Kulkarni}, S.~R., {Djorgovski}, S.~G., {et~al.} 1999, \nat, 401, 453, \dodoi{10.1038/46744}

\bibitem[{{Bobrick} {et~al.}(2022){Bobrick}, {Zenati}, {Perets}, {Davies}, \& {Church}}]{Alexey2022MNRAS}
{Bobrick}, A., {Zenati}, Y., {Perets}, H.~B., {Davies}, M.~B., \& {Church}, R. 2022, \mnras, 510, 3758, \dodoi{10.1093/mnras/stab3574}

\bibitem[{Coulter {et~al.}(2017)Coulter, Foley, Kilpatrick, Drout, Piro, Shappee, Siebert, Simon, Ulloa, Kasen, Madore, Murguia-Berthier, Pan, Prochaska, Ramirez-Ruiz, Rest, \& Rojas-Bravo}]{Coulter2017science}
Coulter, D.~A., Foley, R.~J., Kilpatrick, C.~D., {et~al.} 2017, Science, 358, 1556, \dodoi{10.1126/science.aap9811}

\bibitem[{{Dai} {et~al.}(2023){Dai}, {Guo}, {Zhang}, {Liu}, \& {Wang}}]{Daicuiyuan2023arXiv}
{Dai}, C.-Y., {Guo}, C.-L., {Zhang}, H.-M., {Liu}, R.-Y., \& {Wang}, X.-Y. 2023, arXiv e-prints, arXiv:2312.01074, \dodoi{10.48550/arXiv.2312.01074}

\bibitem[{{Della Valle} {et~al.}(2006){Della Valle}, {Chincarini}, {Panagia}, {Tagliaferri}, {Malesani}, {Testa}, {Fugazza}, {Campana}, {Covino}, {Mangano}, {Antonelli}, {D'Avanzo}, {Hurley}, {Mirabel}, {Pellizza}, {Piranomonte}, \& {Stella}}]{Della2006Natur}
{Della Valle}, M., {Chincarini}, G., {Panagia}, N., {et~al.} 2006, \nat, 444, 1050, \dodoi{10.1038/nature05374}

\bibitem[{{Dichiara} {et~al.}(2023){Dichiara}, {Tsang}, {Troja}, {Neill}, {Norris}, \& {Yang}}]{Dichiara2023ApJ}
{Dichiara}, S., {Tsang}, D., {Troja}, E., {et~al.} 2023, \apjl, 954, L29, \dodoi{10.3847/2041-8213/acf21d}

\bibitem[{{Duncan} \& {Thompson}(1992)}]{Duncan1992ApJ}
{Duncan}, R.~C., \& {Thompson}, C. 1992, \apjl, 392, L9, \dodoi{10.1086/186413}

\bibitem[{{Eichler} {et~al.}(1989){Eichler}, {Livio}, {Piran}, \& {Schramm}}]{Eichler1989Natur}
{Eichler}, D., {Livio}, M., {Piran}, T., \& {Schramm}, D.~N. 1989, \nat, 340, 126, \dodoi{10.1038/340126a0}

\bibitem[{{Fern{\'a}ndez} {et~al.}(2019){Fern{\'a}ndez}, {Margalit}, \& {Metzger}}]{fernndez2019MNRAS}
{Fern{\'a}ndez}, R., {Margalit}, B., \& {Metzger}, B.~D. 2019, \mnras, 488, 259, \dodoi{10.1093/mnras/stz1701}

\bibitem[{{Foreman-Mackey} {et~al.}(2013){Foreman-Mackey}, {Hogg}, {Lang}, \& {Goodman}}]{Foreman-Mackey2013PASP}
{Foreman-Mackey}, D., {Hogg}, D.~W., {Lang}, D., \& {Goodman}, J. 2013, \pasp, 125, 306, \dodoi{10.1086/670067}

\bibitem[{{Freiburghaus} {et~al.}(1999){Freiburghaus}, {Rosswog}, \& {Thielemann}}]{Freiburghaus1999ApJ}
{Freiburghaus}, C., {Rosswog}, S., \& {Thielemann}, F.~K. 1999, \apjl, 525, L121, \dodoi{10.1086/312343}

\bibitem[{{Galama} {et~al.}(1998){Galama}, {Vreeswijk}, {van Paradijs}, {Kouveliotou}, {Augusteijn}, {B{\"o}hnhardt}, {Brewer}, {Doublier}, {Gonzalez}, {Leibundgut}, {Lidman}, {Hainaut}, {Patat}, {Heise}, {in't Zand}, {Hurley}, {Groot}, {Strom}, {Mazzali}, {Iwamoto}, {Nomoto}, {Umeda}, {Nakamura}, {Young}, {Suzuki}, {Shigeyama}, {Koshut}, {Kippen}, {Robinson}, {de Wildt}, {Wijers}, {Tanvir}, {Greiner}, {Pian}, {Palazzi}, {Frontera}, {Masetti}, {Nicastro}, {Feroci}, {Costa}, {Piro}, {Peterson}, {Tinney}, {Boyle}, {Cannon}, {Stathakis}, {Sadler}, {Begam}, \& {Ianna}}]{Galama1998Natur}
{Galama}, T.~J., {Vreeswijk}, P.~M., {van Paradijs}, J., {et~al.} 1998, \nat, 395, 670, \dodoi{10.1038/27150}

\bibitem[{{Gillanders} {et~al.}(2020){Gillanders}, {Sim}, \& {Smartt}}]{Gillanders2020MNRAS}
{Gillanders}, J.~H., {Sim}, S.~A., \& {Smartt}, S.~J. 2020, \mnras, 497, 246, \dodoi{10.1093/mnras/staa1822}

\bibitem[{{Gillanders} {et~al.}(2023){Gillanders}, {Troja}, {Fryer}, {Ristic}, {O'Connor}, {Fontes}, {Yang}, {Domoto}, {Rahmouni}, {Tanaka}, {Fox}, \& {Dichiara}}]{Gillanders2023arXiv}
{Gillanders}, J.~H., {Troja}, E., {Fryer}, C.~L., {et~al.} 2023, arXiv e-prints, arXiv:2308.00633, \dodoi{10.48550/arXiv.2308.00633}

\bibitem[{{Goldstein} {et~al.}(2017){Goldstein}, {Veres}, {Burns}, {Briggs}, {Hamburg}, {Kocevski}, {Wilson-Hodge}, {Preece}, {Poolakkil}, {Roberts}, {Hui}, {Connaughton}, {Racusin}, {von Kienlin}, {Dal Canton}, {Christensen}, {Littenberg}, {Siellez}, {Blackburn}, {Broida}, {Bissaldi}, {Cleveland}, {Gibby}, {Giles}, {Kippen}, {McBreen}, {McEnery}, {Meegan}, {Paciesas}, \& {Stanbro}}]{Goldstein2017ApJ...848L..14G}
{Goldstein}, A., {Veres}, P., {Burns}, E., {et~al.} 2017, \apjl, 848, L14, \dodoi{10.3847/2041-8213/aa8f41}

\bibitem[{{Gottlieb} {et~al.}(2023{\natexlab{a}}){Gottlieb}, {Metzger}, {Quataert}, {Issa}, {Martineau}, {Foucart}, {Duez}, {Kidder}, {Pfeiffer}, \& {Scheel}}]{Ore2023arXivb}
{Gottlieb}, O., {Metzger}, B.~D., {Quataert}, E., {et~al.} 2023{\natexlab{a}}, \apjl, 958, L33, \dodoi{10.3847/2041-8213/ad096e}

\bibitem[{{Gottlieb} {et~al.}(2023{\natexlab{b}}){Gottlieb}, {Issa}, {Jacquemin-Ide}, {Liska}, {Foucart}, {Tchekhovskoy}, {Metzger}, {Quataert}, {Perna}, {Kasen}, {Duez}, {Kidder}, {Pfeiffer}, \& {Scheel}}]{Ore2023ApJ}
{Gottlieb}, O., {Issa}, D., {Jacquemin-Ide}, J., {et~al.} 2023{\natexlab{b}}, \apjl, 954, L21, \dodoi{10.3847/2041-8213/aceeff}

\bibitem[{{Hjorth} {et~al.}(2003){Hjorth}, {Sollerman}, {M{\o}ller}, {Fynbo}, {Woosley}, {Kouveliotou}, {Tanvir}, {Greiner}, {Andersen}, {Castro-Tirado}, {Castro Cer{\'o}n}, {Fruchter}, {Gorosabel}, {Jakobsson}, {Kaper}, {Klose}, {Masetti}, {Pedersen}, {Pedersen}, {Pian}, {Palazzi}, {Rhoads}, {Rol}, {van den Heuvel}, {Vreeswijk}, {Watson}, \& {Wijers}}]{Hjorth2003Natur}
{Hjorth}, J., {Sollerman}, J., {M{\o}ller}, P., {et~al.} 2003, \nat, 423, 847, \dodoi{10.1038/nature01750}

\bibitem[{{Hotokezaka} {et~al.}(2013){Hotokezaka}, {Kiuchi}, {Kyutoku}, {Okawa}, {Sekiguchi}, {Shibata}, \& {Taniguchi}}]{Hotokezaka2013PhRvD}
{Hotokezaka}, K., {Kiuchi}, K., {Kyutoku}, K., {et~al.} 2013, \prd, 87, 024001, \dodoi{10.1103/PhysRevD.87.024001}

\bibitem[{{Huang} {et~al.}(1999){Huang}, {Dai}, \& {Lu}}]{Huangyongfeng1999MNRAS}
{Huang}, Y.~F., {Dai}, Z.~G., \& {Lu}, T. 1999, \mnras, 309, 513, \dodoi{10.1046/j.1365-8711.1999.02887.x}

\bibitem[{{Inserra} {et~al.}(2013){Inserra}, {Smartt}, {Jerkstrand}, {Valenti}, {Fraser}, {Wright}, {Smith}, {Chen}, {Kotak}, {Pastorello}, {Nicholl}, {Bresolin}, {Kudritzki}, {Benetti}, {Botticella}, {Burgett}, {Chambers}, {Ergon}, {Flewelling}, {Fynbo}, {Geier}, {Hodapp}, {Howell}, {Huber}, {Kaiser}, {Leloudas}, {Magill}, {Magnier}, {McCrum}, {Metcalfe}, {Price}, {Rest}, {Sollerman}, {Sweeney}, {Taddia}, {Taubenberger}, {Tonry}, {Wainscoat}, {Waters}, \& {Young}}]{Inserra2013ApJ}
{Inserra}, C., {Smartt}, S.~J., {Jerkstrand}, A., {et~al.} 2013, \apj, 770, 128, \dodoi{10.1088/0004-637X/770/2/128}

\bibitem[{{Kalogera} {et~al.}(1998){Kalogera}, {Kolb}, \& {King}}]{Kalogera1998ApJ}
{Kalogera}, V., {Kolb}, U., \& {King}, A.~R. 1998, \apj, 504, 967, \dodoi{10.1086/306119}

\bibitem[{{Kaltenborn} {et~al.}(2023){Kaltenborn}, {Fryer}, {Wollaeger}, {Belczynski}, {Even}, \& {Kouveliotou}}]{Kaltenborn2022arXiv220913061K}
{Kaltenborn}, M. A.~R., {Fryer}, C.~L., {Wollaeger}, R.~T., {et~al.} 2023, \apj, 956, 71, \dodoi{10.3847/1538-4357/acf860}

\bibitem[{{Kang} {et~al.}(2024){Kang}, {Liu}, {Zhu}, {Gao}, {Shao}, {Zhang}, {Sun}, {Yin}, \& {Zhang}}]{kangyacheng2023arXiv}
{Kang}, Y., {Liu}, C., {Zhu}, J.-P., {et~al.} 2024, \mnras, \dodoi{10.1093/mnras/stae340}

\bibitem[{{Kasen} \& {Bildsten}(2010)}]{Kasen2010ApJ}
{Kasen}, D., \& {Bildsten}, L. 2010, \apj, 717, 245, \dodoi{10.1088/0004-637X/717/1/245}

\bibitem[{{Levan} {et~al.}(2023){Levan}, {Gompertz}, {Salafia}, {Bulla}, {Burns}, {Hotokezaka}, {Izzo}, {Lamb}, {Malesani}, {Oates}, {Ravasio}, {Rouco Escorial}, {Schneider}, {Sarin}, {Schulze}, {Tanvir}, {Ackley}, {Anderson}, {Brammer}, {Christensen}, {Dhillon}, {Evans}, {Fausnaugh}, {Fong}, {Fruchter}, {Fryer}, {Fynbo}, {Gaspari}, {Heintz}, {Hjorth}, {Kennea}, {Kennedy}, {Laskar}, {Leloudas}, {Mandel}, {Martin-Carrillo}, {Metzger}, {Nicholl}, {Nugent}, {Palmerio}, {Pugliese}, {Rastinejad}, {Rhodes}, {Rossi}, {Smartt}, {Stevance}, {Tohuvavohu}, {van der Horst}, {Vergani}, {Watson}, {Barclay}, {Bhirombhakdi}, {Breedt}, {Breeveld}, {Brown}, {Campana}, {Chrimes}, {D'Avanzo}, {D'Elia}, {De Pasquale}, {Dyer}, {Galloway}, {Garbutt}, {Green}, {Hartmann}, {Jakobsson}, {Kerry}, {Langeroodi}, {Leung}, {Littlefair}, {Munday}, {O'Brien}, {Parsons}, {Pelisoli}, {Saccardi}, {Sahman}, {Salvaterra}, {Sbarufatti}, {Steeghs}, {Tagliaferri}, {Th{\"o}ne}, {de Ugarte Postigo}, \& {Kann}}]{Levan2023}
{Levan}, A., {Gompertz}, B.~P., {Salafia}, O.~S., {et~al.} 2023, Nature, \dodoi{10.1038/s41586-023-06759-1}

\bibitem[{{Li} \& {Paczy{\'n}ski}(1998)}]{Lilixin1998ApJ}
{Li}, L.-X., \& {Paczy{\'n}ski}, B. 1998, \apjl, 507, L59, \dodoi{10.1086/311680}

\bibitem[{{Li} {et~al.}(2018){Li}, {Liu}, {Yu}, \& {Zhang}}]{Li&Yu2018ApJ}
{Li}, S.-Z., {Liu}, L.-D., {Yu}, Y.-W., \& {Zhang}, B. 2018, \apjl, 861, L12, \dodoi{10.3847/2041-8213/aace61}

\bibitem[{{Liu} {et~al.}(2021){Liu}, {Gao}, {Wang}, \& {Yang}}]{Liangduan2021ApJ}
{Liu}, L.-D., {Gao}, H., {Wang}, X.-F., \& {Yang}, S. 2021, \apj, 911, 142, \dodoi{10.3847/1538-4357/abf042}

\bibitem[{{Maeda} {et~al.}(2003){Maeda}, {Mazzali}, {Deng}, {Nomoto}, {Yoshii}, {Tomita}, \& {Kobayashi}}]{Maeda2003ApJ}
{Maeda}, K., {Mazzali}, P.~A., {Deng}, J., {et~al.} 2003, \apj, 593, 931, \dodoi{10.1086/376591}

\bibitem[{{Martin} {et~al.}(2015){Martin}, {Perego}, {Arcones}, {Thielemann}, {Korobkin}, \& {Rosswog}}]{Martin2015ApJ}
{Martin}, D., {Perego}, A., {Arcones}, A., {et~al.} 2015, \apj, 813, 2, \dodoi{10.1088/0004-637X/813/1/2}

\bibitem[{{Meng} {et~al.}(2023){Meng}, {Wang}, \& {Liu}}]{Mengyanzhi2023arXiv}
{Meng}, Y.-Z., {Wang}, X.~I., \& {Liu}, Z.-K. 2023, arXiv e-prints, arXiv:2304.00893, \dodoi{10.48550/arXiv.2304.00893}

\bibitem[{{Meszaros} \& {Rees}(1992)}]{Meszaros1992MNRAS}
{Meszaros}, P., \& {Rees}, M.~J. 1992, \mnras, 257, 29P, \dodoi{10.1093/mnras/257.1.29P}

\bibitem[{{Metzger}(2012)}]{Metzger2012MNRAS}
{Metzger}, B.~D. 2012, \mnras, 419, 827, \dodoi{10.1111/j.1365-2966.2011.19747.x}

\bibitem[{{Metzger}(2017)}]{Metzger2017LRR}
---. 2017, Living Reviews in Relativity, 20, 3, \dodoi{10.1007/s41114-017-0006-z}

\bibitem[{{Metzger} \& {Piro}(2014)}]{Metzger2014MNRAS}
{Metzger}, B.~D., \& {Piro}, A.~L. 2014, \mnras, 439, 3916, \dodoi{10.1093/mnras/stu247}

\bibitem[{{Metzger} {et~al.}(2010){Metzger}, {Mart{\'\i}nez-Pinedo}, {Darbha}, {Quataert}, {Arcones}, {Kasen}, {Thomas}, {Nugent}, {Panov}, \& {Zinner}}]{Metzger2010MNRAS}
{Metzger}, B.~D., {Mart{\'\i}nez-Pinedo}, G., {Darbha}, S., {et~al.} 2010, \mnras, 406, 2650, \dodoi{10.1111/j.1365-2966.2010.16864.x}

\bibitem[{{Nagakura} {et~al.}(2014){Nagakura}, {Hotokezaka}, {Sekiguchi}, {Shibata}, \& {Ioka}}]{Nagakura2014ApJ}
{Nagakura}, H., {Hotokezaka}, K., {Sekiguchi}, Y., {Shibata}, M., \& {Ioka}, K. 2014, \apjl, 784, L28, \dodoi{10.1088/2041-8205/784/2/L28}

\bibitem[{{Narayan} {et~al.}(1992){Narayan}, {Paczynski}, \& {Piran}}]{Narayan1992ApJ}
{Narayan}, R., {Paczynski}, B., \& {Piran}, T. 1992, \apjl, 395, L83, \dodoi{10.1086/186493}

\bibitem[{{Nicholl} {et~al.}(2017){Nicholl}, {Guillochon}, \& {Berger}}]{Nicholl2017ApJ}
{Nicholl}, M., {Guillochon}, J., \& {Berger}, E. 2017, \apj, 850, 55, \dodoi{10.3847/1538-4357/aa9334}

\bibitem[{{Nicholl} {et~al.}(2016){Nicholl}, {Berger}, {Smartt}, {Margutti}, {Kamble}, {Alexander}, {Chen}, {Inserra}, {Arcavi}, {Blanchard}, {Cartier}, {Chambers}, {Childress}, {Chornock}, {Cowperthwaite}, {Drout}, {Flewelling}, {Fraser}, {Gal-Yam}, {Galbany}, {Harmanen}, {Holoien}, {Hosseinzadeh}, {Howell}, {Huber}, {Jerkstrand}, {Kankare}, {Kochanek}, {Lin}, {Lunnan}, {Magnier}, {Maguire}, {McCully}, {McDonald}, {Metzger}, {Milisavljevic}, {Mitra}, {Reynolds}, {Saario}, {Shappee}, {Smith}, {Valenti}, {Villar}, {Waters}, \& {Young}}]{Nicholl2016ApJ}
{Nicholl}, M., {Berger}, E., {Smartt}, S.~J., {et~al.} 2016, \apj, 826, 39, \dodoi{10.3847/0004-637X/826/1/39}

\bibitem[{{Nomoto} \& {Kondo}(1991)}]{Nomoto1991ApJ}
{Nomoto}, K., \& {Kondo}, Y. 1991, \apjl, 367, L19, \dodoi{10.1086/185922}

\bibitem[{{Pian} {et~al.}(2017){Pian}, {D'Avanzo}, {Benetti}, {Branchesi}, {Brocato}, {Campana}, {Cappellaro}, {Covino}, {D'Elia}, {Fynbo}, {Getman}, {Ghirlanda}, {Ghisellini}, {Grado}, {Greco}, {Hjorth}, {Kouveliotou}, {Levan}, {Limatola}, {Malesani}, {Mazzali}, {Melandri}, {M{\o}ller}, {Nicastro}, {Palazzi}, {Piranomonte}, {Rossi}, {Salafia}, {Selsing}, {Stratta}, {Tanaka}, {Tanvir}, {Tomasella}, {Watson}, {Yang}, {Amati}, {Antonelli}, {Ascenzi}, {Bernardini}, {Bo{\"e}r}, {Bufano}, {Bulgarelli}, {Capaccioli}, {Casella}, {Castro-Tirado}, {Chassande-Mottin}, {Ciolfi}, {Copperwheat}, {Dadina}, {De Cesare}, {di Paola}, {Fan}, {Gendre}, {Giuffrida}, {Giunta}, {Hunt}, {Israel}, {Jin}, {Kasliwal}, {Klose}, {Lisi}, {Longo}, {Maiorano}, {Mapelli}, {Masetti}, {Nava}, {Patricelli}, {Perley}, {Pescalli}, {Piran}, {Possenti}, {Pulone}, {Razzano}, {Salvaterra}, {Schipani}, {Spera}, {Stamerra}, {Stella}, {Tagliaferri}, {Testa}, {Troja}, {Turatto}, {Vergani}, \& {Vergani}}]{Pian2017Natur}
{Pian}, E., {D'Avanzo}, P., {Benetti}, S., {et~al.} 2017, \nat, 551, 67, \dodoi{10.1038/nature24298}

\bibitem[{{Pinto} \& {Eastman}(2000)}]{Pinto2000ApJ}
{Pinto}, P.~A., \& {Eastman}, R.~G. 2000, \apj, 530, 744, \dodoi{10.1086/308376}

\bibitem[{{Radice} {et~al.}(2018){Radice}, {Perego}, {Hotokezaka}, {Fromm}, {Bernuzzi}, \& {Roberts}}]{Radice2018ApJ}
{Radice}, D., {Perego}, A., {Hotokezaka}, K., {et~al.} 2018, \apj, 869, 130, \dodoi{10.3847/1538-4357/aaf054}

\bibitem[{{Rastinejad} {et~al.}(2022){Rastinejad}, {Gompertz}, {Levan}, {Fong}, {Nicholl}, {Lamb}, {Malesani}, {Nugent}, {Oates}, {Tanvir}, {de Ugarte Postigo}, {Kilpatrick}, {Moore}, {Metzger}, {Ravasio}, {Rossi}, {Schroeder}, {Jencson}, {Sand}, {Smith}, {Ag{\"u}{\'\i} Fern{\'a}ndez}, {Berger}, {Blanchard}, {Chornock}, {Cobb}, {De Pasquale}, {Fynbo}, {Izzo}, {Kann}, {Laskar}, {Marini}, {Paterson}, {Escorial}, {Sears}, \& {Th{\"o}ne}}]{Rastinejad2022Natur}
{Rastinejad}, J.~C., {Gompertz}, B.~P., {Levan}, A.~J., {et~al.} 2022, \nat, 612, 223, \dodoi{10.1038/s41586-022-05390-w}

\bibitem[{{Ryan} {et~al.}(2020){Ryan}, {van Eerten}, {Piro}, \& {Troja}}]{Ryan2020ApJ}
{Ryan}, G., {van Eerten}, H., {Piro}, L., \& {Troja}, E. 2020, \apj, 896, 166, \dodoi{10.3847/1538-4357/ab93cf}

\bibitem[{{Sari} {et~al.}(1998){Sari}, {Piran}, \& {Narayan}}]{Sari1998ApJ}
{Sari}, R., {Piran}, T., \& {Narayan}, R. 1998, \apjl, 497, L17, \dodoi{10.1086/311269}

\bibitem[{{Savchenko} {et~al.}(2017){Savchenko}, {Ferrigno}, {Kuulkers}, {Bazzano}, {Bozzo}, {Brandt}, {Chenevez}, {Courvoisier}, {Diehl}, {Domingo}, {Hanlon}, {Jourdain}, {von Kienlin}, {Laurent}, {Lebrun}, {Lutovinov}, {Martin-Carrillo}, {Mereghetti}, {Natalucci}, {Rodi}, {Roques}, {Sunyaev}, \& {Ubertini}}]{Savchenko2017ApJ}
{Savchenko}, V., {Ferrigno}, C., {Kuulkers}, E., {et~al.} 2017, \apjl, 848, L15, \dodoi{10.3847/2041-8213/aa8f94}

\bibitem[{{Shigeyama} {et~al.}(1987){Shigeyama}, {Nomoto}, {Hashimoto}, \& {Sugimoto}}]{Shigeyama1987Natur_SN1987}
{Shigeyama}, T., {Nomoto}, K., {Hashimoto}, M., \& {Sugimoto}, D. 1987, \nat, 328, 320, \dodoi{10.1038/328320a0}

\bibitem[{{Smartt} {et~al.}(2017){Smartt}, {Chen}, {Jerkstrand}, {Coughlin}, {Kankare}, {Sim}, {Fraser}, {Inserra}, {Maguire}, {Chambers}, {Huber}, {Kr{\"u}hler}, {Leloudas}, {Magee}, {Shingles}, {Smith}, {Young}, {Tonry}, {Kotak}, {Gal-Yam}, {Lyman}, {Homan}, {Agliozzo}, {Anderson}, {Angus}, {Ashall}, {Barbarino}, {Bauer}, {Berton}, {Botticella}, {Bulla}, {Bulger}, {Cannizzaro}, {Cano}, {Cartier}, {Cikota}, {Clark}, {De Cia}, {Della Valle}, {Denneau}, {Dennefeld}, {Dessart}, {Dimitriadis}, {Elias-Rosa}, {Firth}, {Flewelling}, {Fl{\"o}rs}, {Franckowiak}, {Frohmaier}, {Galbany}, {Gonz{\'a}lez-Gait{\'a}n}, {Greiner}, {Gromadzki}, {Guelbenzu}, {Guti{\'e}rrez}, {Hamanowicz}, {Hanlon}, {Harmanen}, {Heintz}, {Heinze}, {Hernandez}, {Hodgkin}, {Hook}, {Izzo}, {James}, {Jonker}, {Kerzendorf}, {Klose}, {Kostrzewa-Rutkowska}, {Kowalski}, {Kromer}, {Kuncarayakti}, {Lawrence}, {Lowe}, {Magnier}, {Manulis}, {Martin-Carrillo}, {Mattila}, {McBrien}, {M{\"u}ller}, {Nordin}, {O'Neill}, {Onori}, {Palmerio}, {Pastorello},
  {Patat}, {Pignata}, {Podsiadlowski}, {Pumo}, {Prentice}, {Rau}, {Razza}, {Rest}, {Reynolds}, {Roy}, {Ruiter}, {Rybicki}, {Salmon}, {Schady}, {Schultz}, {Schweyer}, {Seitenzahl}, {Smith}, {Sollerman}, {Stalder}, {Stubbs}, {Sullivan}, {Szegedi}, {Taddia}, {Taubenberger}, {Terreran}, {van Soelen}, {Vos}, {Wainscoat}, {Walton}, {Waters}, {Weiland}, {Willman}, {Wiseman}, {Wright}, {Wyrzykowski}, \& {Yaron}}]{Smartt2017Natur}
{Smartt}, S.~J., {Chen}, T.~W., {Jerkstrand}, A., {et~al.} 2017, \nat, 551, 75, \dodoi{10.1038/nature24303}

\bibitem[{{Stanek} {et~al.}(2003){Stanek}, {Matheson}, {Garnavich}, {Martini}, {Berlind}, {Caldwell}, {Challis}, {Brown}, {Schild}, {Krisciunas}, {Calkins}, {Lee}, {Hathi}, {Jansen}, {Windhorst}, {Echevarria}, {Eisenstein}, {Pindor}, {Olszewski}, {Harding}, {Holland}, \& {Bersier}}]{Stanek2003ApJ}
{Stanek}, K.~Z., {Matheson}, T., {Garnavich}, P.~M., {et~al.} 2003, \apjl, 591, L17, \dodoi{10.1086/376976}

\bibitem[{{Sun} {et~al.}(2023){Sun}, {Wang}, {Yang}, {Zhang}, {Xiong}, {Yin}, {Liu}, {Li}, {Xue}, {Yan}, {Zhang}, {Tan}, {Pan}, {Liu}, {Cheng}, {Zhang}, {Hu}, {Zheng}, {An}, {Cai}, {Hu}, {Jin}, {Li}, {Li}, {Liu}, {Liu}, {Peng}, {Song}, {Sun}, {Sun}, {Wang}, {Wen}, {Xiao}, {Yi}, {Zhang}, {Zhang}, {Zhang}, {Zhang}, {Zhao}, {Zheng}, {Ling}, {Zhang}, {Yuan}, \& {Zhang}}]{sunhui2023arXiv}
{Sun}, H., {Wang}, C.~W., {Yang}, J., {et~al.} 2023, arXiv e-prints, arXiv:2307.05689, \dodoi{10.48550/arXiv.2307.05689}

\bibitem[{{Sutherland} \& {Wheeler}(1984)}]{Sutherland1984ApJ}
{Sutherland}, P.~G., \& {Wheeler}, J.~C. 1984, \apj, 280, 282, \dodoi{10.1086/161995}

\bibitem[{{Taam} \& {van den Heuvel}(1986)}]{Taam1986ApJ}
{Taam}, R.~E., \& {van den Heuvel}, E.~P.~J. 1986, \apj, 305, 235, \dodoi{10.1086/164243}

\bibitem[{{Troja} {et~al.}(2022){Troja}, {Fryer}, {O'Connor}, {Ryan}, {Dichiara}, {Kumar}, {Ito}, {Gupta}, {Wollaeger}, {Norris}, {Kawai}, {Butler}, {Aryan}, {Misra}, {Hosokawa}, {Murata}, {Niwano}, {Pandey}, {Kutyrev}, {van Eerten}, {Chase}, {Hu}, {Caballero-Garcia}, \& {Castro-Tirado}}]{Troja2022Natur}
{Troja}, E., {Fryer}, C.~L., {O'Connor}, B., {et~al.} 2022, \nat, 612, 228, \dodoi{10.1038/s41586-022-05327-3}

\bibitem[{{Usov}(1992)}]{Usov1992Natur}
{Usov}, V.~V. 1992, \nat, 357, 472, \dodoi{10.1038/357472a0}

\bibitem[{{Wang} {et~al.}(2023){Wang}, {Xia}, {Zheng}, {Ren}, \& {Fan}}]{Wangyun2023ApJ}
{Wang}, Y., {Xia}, Z.-Q., {Zheng}, T.-C., {Ren}, J., \& {Fan}, Y.-Z. 2023, \apjl, 953, L8, \dodoi{10.3847/2041-8213/ace7d4}

\bibitem[{{Woosley} \& {Bloom}(2006)}]{Woosley2006ARA&A}
{Woosley}, S.~E., \& {Bloom}, J.~S. 2006, \araa, 44, 507, \dodoi{10.1146/annurev.astro.43.072103.150558}

\bibitem[{{Yang} {et~al.}(2015){Yang}, {Jin}, {Li}, {Covino}, {Zheng}, {Hotokezaka}, {Fan}, {Piran}, \& {Wei}}]{Yangbin2015NatCo}
{Yang}, B., {Jin}, Z.-P., {Li}, X., {et~al.} 2015, Nature Communications, 6, 7323, \dodoi{10.1038/ncomms8323}

\bibitem[{{Yang} {et~al.}(2022){Yang}, {Ai}, {Zhang}, {Zhang}, {Liu}, {Wang}, {Yang}, {Yin}, {Li}, \& {L{\"u}}}]{Yangjun2022Natur}
{Yang}, J., {Ai}, S., {Zhang}, B.-B., {et~al.} 2022, \nat, 612, 232, \dodoi{10.1038/s41586-022-05403-8}

\bibitem[{{Yang} {et~al.}(2023){Yang}, {Troja}, {O'Connor}, {Fryer}, {Im}, {Durbak}, {Paek}, {Ricci}, {De Bom}, {Gillanders}, {Castro-Tirado}, {Peng}, {Dichiara}, {Ryan}, {van Eerten}, {Dai}, {Chang}, {Choi}, {De}, {Hu}, {Kilpatrick}, {Kutyrev}, {Jeong}, {Lee}, {Makler}, {Navarete}, \& {P{\'e}rez-Garc{\'\i}a}}]{yangyuhan2023arXiv}
{Yang}, Y.-H., {Troja}, E., {O'Connor}, B., {et~al.} 2023, arXiv e-prints, arXiv:2308.00638, \dodoi{10.48550/arXiv.2308.00638}

\bibitem[{{Yin} {et~al.}(2023){Yin}, {Zhang}, {Sun}, {Yang}, {Kang}, {Shao}, {Yang}, \& {Zhang}}]{Yinyihan2023ApJ}
{Yin}, Y.-H.~I., {Zhang}, B.-B., {Sun}, H., {et~al.} 2023, \apjl, 954, L17, \dodoi{10.3847/2041-8213/acf04a}

\bibitem[{{Yoon} {et~al.}(2007){Yoon}, {Podsiadlowski}, \& {Rosswog}}]{Yoon2007MNRAS}
{Yoon}, S.~C., {Podsiadlowski}, P., \& {Rosswog}, S. 2007, \mnras, 380, 933, \dodoi{10.1111/j.1365-2966.2007.12161.x}

\bibitem[{{Yu} {et~al.}(2019{\natexlab{a}}){Yu}, {Chen}, \& {Li}}]{Yu&chen&li2019ApJ}
{Yu}, Y.-W., {Chen}, A., \& {Li}, X.-D. 2019{\natexlab{a}}, \apjl, 877, L21, \dodoi{10.3847/2041-8213/ab1f85}

\bibitem[{{Yu} {et~al.}(2019{\natexlab{b}}){Yu}, {Chen}, \& {Wang}}]{Yu&chen&wang2019ApJ}
{Yu}, Y.-W., {Chen}, A., \& {Wang}, B. 2019{\natexlab{b}}, \apjl, 870, L23, \dodoi{10.3847/2041-8213/aaf960}

\bibitem[{{Yu} {et~al.}(2015{\natexlab{a}}){Yu}, {Li}, \& {Dai}}]{Yu2015ApJ}
{Yu}, Y.-W., {Li}, S.-Z., \& {Dai}, Z.-G. 2015{\natexlab{a}}, \apjl, 806, L6, \dodoi{10.1088/2041-8205/806/1/L6}

\bibitem[{{Yu} {et~al.}(2015{\natexlab{b}}){Yu}, {Li}, \& {Dai}}]{Yu&Li2015ApJ}
---. 2015{\natexlab{b}}, \apjl, 806, L6, \dodoi{10.1088/2041-8205/806/1/L6}

\bibitem[{{Yu} {et~al.}(2018){Yu}, {Liu}, \& {Dai}}]{Yu_yunwei2018ApJ}
{Yu}, Y.-W., {Liu}, L.-D., \& {Dai}, Z.-G. 2018, \apj, 861, 114, \dodoi{10.3847/1538-4357/aac6e5}

\bibitem[{{Yu} {et~al.}(2013){Yu}, {Zhang}, \& {Gao}}]{Yunwei_yu2013ApJ}
{Yu}, Y.-W., {Zhang}, B., \& {Gao}, H. 2013, \apjl, 776, L40, \dodoi{10.1088/2041-8205/776/2/L40}

\bibitem[{{Yu} {et~al.}(2017){Yu}, {Zhu}, {Li}, {L{\"u}}, \& {Zou}}]{Yu&Zhu2017ApJ}
{Yu}, Y.-W., {Zhu}, J.-P., {Li}, S.-Z., {L{\"u}}, H.-J., \& {Zou}, Y.-C. 2017, \apj, 840, 12, \dodoi{10.3847/1538-4357/aa6c27}

\bibitem[{{Zenati} {et~al.}(2020){Zenati}, {Bobrick}, \& {Perets}}]{Zenati2020MNRAS}
{Zenati}, Y., {Bobrick}, A., \& {Perets}, H.~B. 2020, \mnras, 493, 3956, \dodoi{10.1093/mnras/staa507}

\bibitem[{{Zenati} {et~al.}(2019){Zenati}, {Perets}, \& {Toonen}}]{Zenati2019MNRAS}
{Zenati}, Y., {Perets}, H.~B., \& {Toonen}, S. 2019, \mnras, 486, 1805, \dodoi{10.1093/mnras/stz316}

\bibitem[{{Zhang}(2018)}]{zhangbing2018pgrb.book}
{Zhang}, B. 2018, {The Physics of Gamma-Ray Bursts}, \dodoi{10.1017/9781139226530}

\bibitem[{{Zhong} {et~al.}(2023){Zhong}, {Li}, \& {Dai}}]{Zhongshuqing2023ApJ}
{Zhong}, S.-Q., {Li}, L., \& {Dai}, Z.-G. 2023, \apjl, 947, L21, \dodoi{10.3847/2041-8213/acca83}

\bibitem[{{Zhu} {et~al.}(2022){Zhu}, {Wang}, {Sun}, {Yang}, {Li}, {Hu}, {Qin}, \& {Wu}}]{Jinpingzhu2022ApJ}
{Zhu}, J.-P., {Wang}, X.~I., {Sun}, H., {et~al.} 2022, \apjl, 936, L10, \dodoi{10.3847/2041-8213/ac85ad}

\end{thebibliography}

\end{CJK*}
\end{document}